\documentclass[aps,preprintnumbers,superscriptaddress,amsmath,amssymb,floatfix,twocolumn]{revtex4}
\usepackage{graphicx}
\usepackage{euscript}
\usepackage{oldgerm}
\usepackage{subfigure}
\usepackage{color}
\usepackage{txfonts}
\usepackage{amssymb}
\usepackage{amsmath}
\begin{document}
\title{Swarm behavior of self-propelled rods and swimming flagella}
\author{Yingzi Yang}
\affiliation{Theoretical Soft Matter and Biophysics Group, Institut
f{\"u}r Festk{\"o}rperforschung, Forschungszentrum J{\"u}lich,
D-52425 J{\"u}lich, Germany}
\author{Vincent Marceau}
\affiliation{Theoretical Soft Matter and Biophysics Group, Institut
f{\"u}r Festk{\"o}rperforschung, Forschungszentrum J{\"u}lich,
D-52425 J{\"u}lich, Germany}
\affiliation{D{\'e}partement de physique, de g{\'e}nie physique et d'optique,
Universit{\'e} Laval, Qu{\'e}bec, Qu{\'e}bec, Canada G1V 0A6}
\author{Gerhard Gompper}
\affiliation{Theoretical Soft Matter and Biophysics Group, Institut
f{\"u}r Festk{\"o}rperforschung, Forschungszentrum J{\"u}lich,
D-52425 J{\"u}lich, Germany}
\date{\today}
\begin{abstract}
Systems of self-propelled particles are known for their tendency to
aggregate and to display swarm behavior. We investigate two model
systems, self-propelled rods interacting via volume exclusion, and
sinusoidally-beating flagella embedded in a fluid with hydrodynamic
interactions. In the flagella system, beating frequencies are
Gaussian distributed with a non-zero average. These systems are
studied by Brownian-dynamics simulations and by mesoscale
hydrodynamics simulations, respectively. The clustering behavior is
analyzed as the particle density and the environmental or internal
noise are varied. By distinguishing three types of cluster-size
probability density functions, we obtain a phase diagram of
different swarm behaviors. The properties of clusters, such as their
configuration, lifetime and average size are analyzed. We find that
the swarm behavior of the two systems, characterized by several
effective power laws, is very similar. However, a more careful
analysis reveals several differences. Clusters of self-propelled
rods form due to partially blocked forward motion, and are therefore
typically wedge-shaped. At higher rod density and low
noise, a giant mobile cluster appears, in which most rods are mostly
oriented towards the center. In contrast, flagella become
hydrodynamically synchronized and attract each other; their clusters
are therefore more elongated. Furthermore, the lifetime of flagella
clusters decays more quickly with cluster size than of rod clusters.
\end{abstract}
\maketitle

%
%
\section{Introduction}
\label{sec:intro}

Systems of self-propelled particles (SPP), which exhibit
an interaction mechanism that favors velocity alignment of
neighboring particles, often display collective behaviors like swarming
and clustering. There are many examples for this swarming behavior,
ranging from systems of microscopic particles (sperm, bacteria, nano-rods)
to systems of macroscopic objects (birds, fish).

Since the pioneering simulation work of Vicsek et al.
\cite{Vicsek1995}, SPP systems have attracted a lot of interest at
the theoretical \cite{Peruani2008, Ramaswamy2003,
Toner1995, Simha2002, Baskaran2008, Baskaran2008b,Bertin2009} and
computational \cite{Szabo2006, Chate2008, Gregoire2004, Huepe2004,
D'Orsogna2006, Aldana2007,Ginelli2010} level. Typically, in
simulation models of swarm behavior, point-like
agents move with an imposed non-zero velocity and tend to align
their direction of motion with others in a prescribed neighborhood
\cite{Vicsek1995, Chate2008, Gregoire2004, Huepe2004}. Although the
alignment mechanism may differ from one model to the other, the
basic properties of swarm behavior are quite universal
\cite{Huepe2008}. Upon variation of parameters such as
particle density, particle velocity, or environmental noise, the
system can undergo a transition from a disordered state, where
the average total velocity or orientation vanishes,
to a nematically ordered state. Near the transition point,
the cluster-size probability density function is characterized by a
power-law decay \cite{Huepe2004,Huepe2008}. For
intermediate densities, phase separation into regions of different
density and band formation has been found \cite{Ginelli2010}.

Self-propelled motion is common in biological systems at micro- or
mesoscopic length scales, such as suspensions of bacteria, like
\emph{E. coli} \cite{Wu2000} and \emph{Bacillus subtilis}
\cite{Ben-Jacob2000,Dombrowski2004,Sokolov2007}, or
tissue cells (keratocytes) \cite{Szabo2006},
whose sizes are all on the micrometers scale. A
special class of biological systems are rod-like self-propelled
particles (rSPP), for example myxobacteria (approximately $10\mu m$
long) \cite{Igoshin2004,Kuner1982}. When starved, myxobacteria are
elongated to an average aspect ratio of approximately 1:7, glide on
a substrate along their long axis and undergo a process of
alignment, rippling, streaming and aggregation that culminates in a
three-dimensional fruiting body.
A model, which takes into account the exchange of a morphogen
during cell-cell contact and a preferred cell motion in the direction
of largest morphogen concentration, has been designed to describe
the streaming and two-stage aggregation of myxobacteria \cite{Alber2004}.

Sperm (with a length of about $50 \mu m$)
\cite{Taylor1951,Gray1957} and nematodes \cite{Gray1964}
(about $1 mm$ long) employ a sinusoidal undulation of their slender
bodies to push the fluid backwards and to propel themselves forward.
Large train-like
clusters of wood mouse sperm \cite{Moore2002, Immler2007} are
believed to result in greater thrust forces to move more efficiently
through a highly viscous environment.
The wood mouse
sperm has a hook-like structure at its head, by which it can be hitched
to the mid-part or the tail of a
neighboring cell for robust cooperation. However, nematodes
which do not have hook structures, also display a
pronounced tendency to adhere to each other in a film of water, to
form assemblies consisting of many organisms, and to exhibit a
striking co-ordinated movement \cite{Gray1964}.
Also, sea urchin
sperm organize into a hexagonal pattern of rotating vortices at
surfaces \cite{Riedel2005}.

A nice physical realization of
self-propelled rods (SPR) are bimetallic
nano-rods consisting of long Pt and Au segments \cite{Paxton2004}.
The rods, about $300nm$ in diameter and $2\mu m$ long, move
autonomously in an aqueous hydrogen peroxide solutions by catalyzing
the formation of oxygen at the Pt end. They move predominantly in
the direction of the Pt end, with a velocity depending on the
concentration of hydrogen peroxide. When a gradient of the hydrogen
peroxide concentration is imposed, the rods exhibit directed motion
towards regions of higher concentrations through active diffusion
\cite{Hong2007}.

A related system is a fluidized monolayer of macroscopic rods in the
nematic liquid crystalline phase \cite{Narayan2007}. The rods
confined between two hard walls are energized by an
external vertical vibration, and gain kinetic energy through
frequent collisions with the floor and the ceiling of the container.
Long-lived giant number fluctuations are found, which shows that
simple contact can give rise to flocking, coherent swirling motion
and large-scale inhomogeneities \cite{Kudrolli2008}. However, in
this experiment, the rods do not have a preferred direction of
motion.

All of these examples of self-propelled particles employ different
propulsion mechanisms and have different interactions.
However, their swarm behavior, such as
flocking, streaming and clustering, is surprisingly
similar. The common characteristic of these systems
is their
rod-like structures and their quasi-two-dimensional active motion.
Myxobacteria glide on surfaces \cite{Kuner1982}, while sperm and
nematodes gather at substrates
\cite{Rothschild1963,Gray1964,Riedel2005}. In
suspensions of rod-like particles in thermal equilibrium, volume
exclusion favors the alignment of rods.  At high densities, it
stabilizes a nematic state characterized by long-range orientational
order \cite{Kayser1978}.

While constant-velocity polar point particles
interacting locally by nematic alignment in the presence of noise
have been studied intensively in recent years \cite{Vicsek1995,
Peruani2008, Ramaswamy2003, Toner1995, Simha2002, Baskaran2008,
Baskaran2008b, Bertin2009, Szabo2006, Chate2008, Gregoire2004,
Huepe2004, D'Orsogna2006, Aldana2007,Ginelli2010}, much less is
known theoretically about the behavior of elongated particles with
volume exclusion, or about the collective behavior of swimmers with
hydrodynamic interactions. Previous simulation studies of
self-propelled rods (SPR) in two dimensions show that self-propelled
motion enhances the tendency for nematic ordering
\cite{Kraikivski2006}, as well as aggregation and clustering
\cite{Peruani2006}. Also, rods have an increased probability to be
located near surfaces (depending on their velocity, length and
thermal noise) \cite{Elgeti2009} and form hedgehog-like clusters at
surfaces \cite{Wensink2008}. In Ref.~\cite{Peruani2006}, two regimes
of clustering have be distinguished by their unimodal
or bimodal weighted cluster-size distribution functions;
however, the system contained a relatively small number of particles
compared to those employed in simulation studies of swarming of
SPPs. Continuum equations for the description of SPR systems have
been derived recently within a mean-field approximation
\cite{Baskaran2008,Baskaran2008b}. This theory predicts that
hard-core interactions are  insufficient to generate a
macroscopically polarized state, because they cannot distinguish the
two ends of a rod, and makes interesting predictions for the
fluctuations in the nematic and isotropic state (such as a crossover
from diffusive to propagating density fluctuations). However, the
mean-field approximation of volume exclusion has the limitation of
omitting correlation effects, and thus works best for slowly varying
density distributions.

In addition, hydrodynamic interactions between rSPP
have so far been largely neglected. These interactions depend on the
type of self-propulsion, where ``pullers" repel  and ``pushers"
attract each other \cite{Ishikawa2009,Lauga2009}. Nematic
suspensions of swimming rod-like pushers are found to be unstable at
long wavelengths as a result of hydrodynamic fluctuations
\cite{Saintillan2007}. For sperm and flagella, it has been shown theoretically
that the hydrodynamic coupling synchronizes the phases of their
sinusoidal beating tails \cite{Taylor1951, Yang2008, Elfring2009}.
Also, the hydrodynamic interaction between these microswimmers
implies attraction and cluster formation \cite{Yang2008}; similarly,
it makes an essential contribution to the capturing of sperm near
walls \cite{Elgeti2010}. However, the relative importance of
directed self-propulsion, particle shape, volume exclusion, and
hydrodynamic interactions to the emergence of swarm behavior remains
unclear.

In this paper, we employ a model of hard rods with strict
volume exclusions and simulate large systems containing at least
1000 particles. We focus on rSPP systems at a density
below the isotropic-nematic transition of Brownian
rods. We employ a model consisting of rigid SPR performing an
overdamped translational motion in two dimensions, and analyze the
resulting cluster-size probability density distribution, cluster
configurations and lifetimes.  Three types of
cluster-size probability density distribution functions allow to
distinguish three different states, and to construct a phase diagram
as a function of particle density and environmental noise. As a
special case of rSPP with an explicit propulsion mechanism, we
investigate a suspension of flagella, which move by sinusoidal
beating of their body in a two-dimensional fluid. The motion of the
surrounding fluid is described by particle-based mesoscopic
simulation method called multi-particle collision dynamics (MPC)
\cite{Kapral2008,Gompper2009}. This method has been shown to capture
the full hydrodynamics and flow behavior of complex fluids over a
wide range of Reynolds numbers very well \cite{Ripoll2004}. By
comparing the results for SPR and flagella, we elucidate the contribution
of hydrodynamic interactions to the swarm behavior.

This paper is organized as follows. Section~\ref{sec:model} gives a brief
description of our models and simulation methods. We analyze the
collective behavior of SPR systems in Sec.~\ref{sec:swarming_rods}.
In Sec.~\ref{sec:swarming_flagella}, we
study the swarm behavior of flagella, and compare the results obtained
with both models. The influence of hydrodynamic interactions and
the flagellar beat on the swarm behavior are discussed. We
summarize our main conclusions in Sec.~\ref{sec:conclusions}.

%
%
\section{Models and Simulation Techniques}
\label{sec:model}

\subsection{Self-Propelled Rods}
\label{sec:model_rods}

\begin{figure}
\subfigure[]{\resizebox{4cm}{!}{\includegraphics{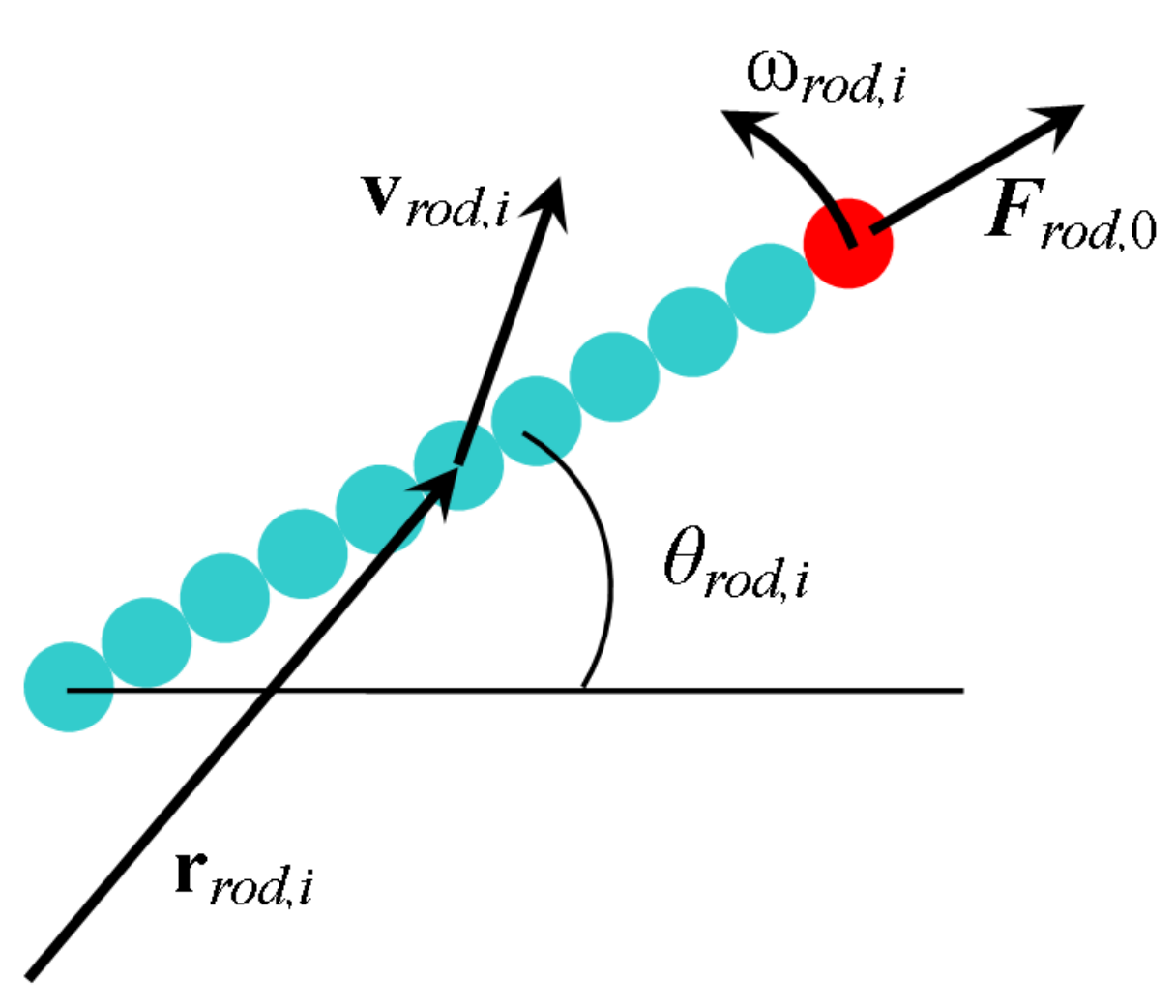}}}
\subfigure[]{\resizebox{4cm}{!}{\includegraphics{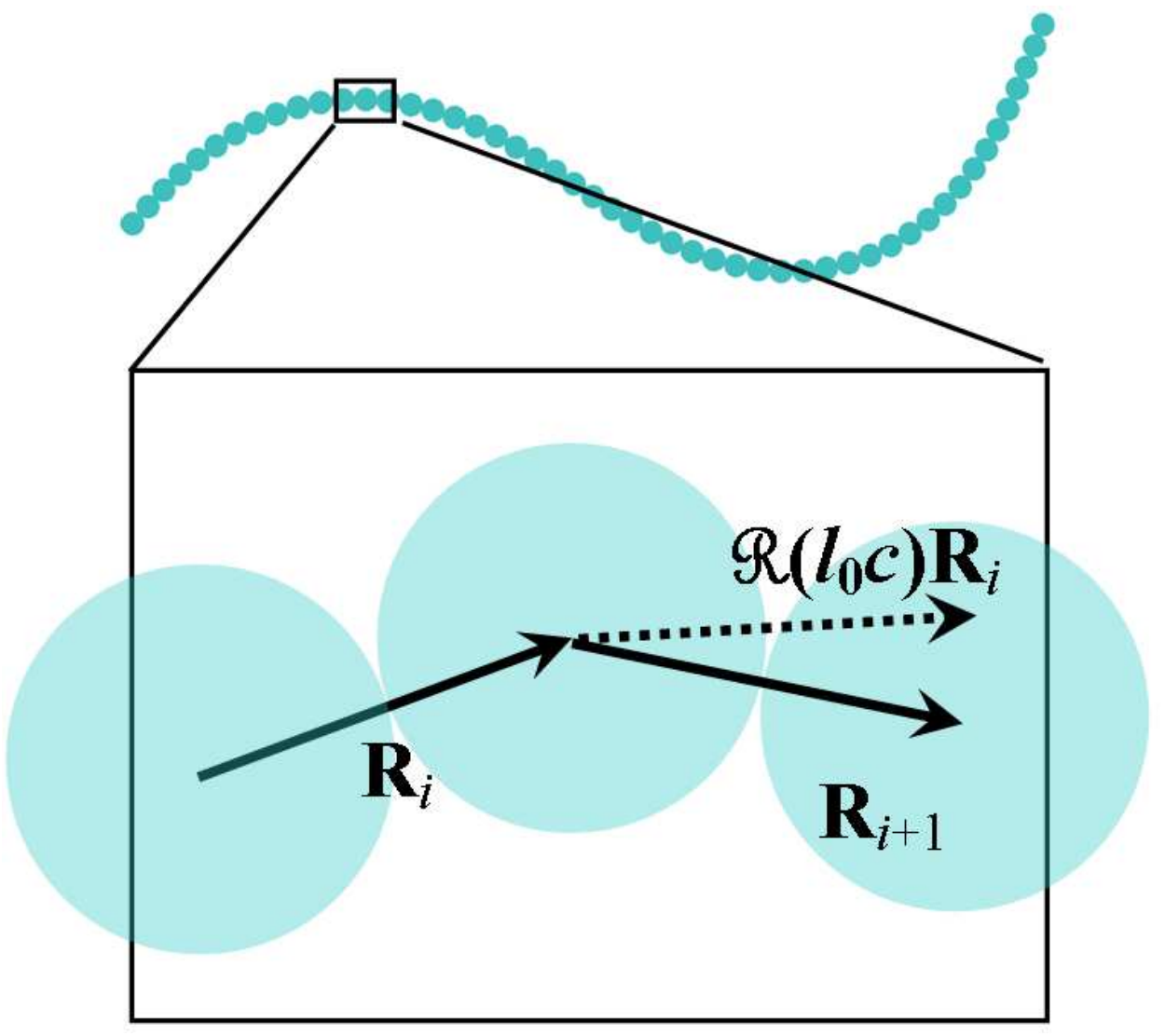}}}
\caption{(Color online) (a) Model of a self-propelled rod, and the
coordinates used in two dimensions. The rod is discretized into
$n_{rod}=L_{rod}/l_b$ beads for the calculation the volume-exclusion
interaction. (b) Model of a flagellum in two dimensions. }
\label{model}
\end{figure}

We consider a system of $N_{rod}$ rods of length $L_{rod}$ in a
two-dimensional simulation box of size $L_x\times L_y$. Each rod is
characterized by an orientation angle $\theta_{rod,i}$
with respect to the $x$-axis, a center-of-mass position
${\bf r}_{rod,i}$, a center-of-mass velocity ${\bf v}_{rod,i}$ and an
angular velocity $\omega_{rod,i}$ around its center of mass (see
Fig.~\ref{model}a). The rods move ballistically according to their
velocities,
\begin{eqnarray}
  {\bf r}_{rod,i}(t+\Delta t_{rod}) &=& {\bf r}_{rod,i}(t) +
                   {\bf v}_{rod,i}(t)\Delta t_{rod}  \ , \\
  \theta_{rod,i}(t+\Delta t_{rod}) &=& \theta_{rod,i}(t) +
                      \omega_{rod,i}(t)\Delta t_{rod} \ ,
\end{eqnarray}
where $\Delta t_{rod}$ is the simulation time step. The particle
velocity can be decomposed into a
parallel and a perpendicular component relative to the rod axis, ${\bf
v}_{rod,i}={\bf v}_{rod,i,\parallel}+{\bf v}_{rod,i,\perp}$.

We consider the rods to be embedded in an overdamped fluid medium where
hydrodynamics can be approximated by an anisotropic friction on the
rod-like particles. The motion is then determined by
\begin{equation}
{\bf v}_{rod,i,\parallel}(t)=
\frac{1}{\gamma_\parallel}\left(\sum_{j\neq i}^{N_{rod}}
    {\bf F}_{ij,\parallel} + \xi_{\parallel}{\bf e}_{\parallel} +
                         F_{rod,0}{\bf e}_{\parallel}\right) \ ,
\end{equation}
\begin{equation}
{\bf v}_{rod,i,\perp}(t)=
\frac{1}{\gamma_\perp}\left(\sum_{j\neq i}^{N_{rod}}
      {\bf F}_{ij,\perp}+\xi_{\perp}{\bf e}_{\perp}\right) \ ,
\end{equation}
\begin{equation}
\omega_{rod,i}(t)=\frac{1}{\gamma_r}\left(\sum_{j\neq i}^{N_{rod}}
M_{ij}+\xi_r\right) \ ,
\end{equation}
where ${\bf e}_\parallel$ and ${\bf e}_\perp$ are the local parallel
and perpendicular unit vectors of the rod orientation. $F_{rod,0}$
is a constant propelling force applied along ${\bf e}_\parallel$.
The friction coefficients are given by
$\gamma_{\perp}=2\gamma_{\parallel}$, $\gamma_{\parallel}=L_{rod}$
and $\gamma_r=\gamma_{\parallel}{L_{rod}}^2/6$. The random forces
$\xi_{\parallel}$, $\xi_{\perp}$ and ${\xi_r}$ are white noises,
which are are determined by their variances $\sigma_{rod}^2L_{rod}$,
$\sigma_{rod}^2L_{rod}$ and $\sigma_{rod}^2L_{rod}^3/12$,
respectively. Finally, ${\bf F}_{ij}$ is the force
generated by volume exclusion between rods $i$ and $j$, and
$M_{ij}$ is the torque generated by ${\bf F}_{ij}$ on rod $i$ in
the reference system of center of mass of rod $i$.

For the calculation of the interactions, each rod is discretized into
$n_{rod} = L_{rod}/l_b$ beads of diameter $l_b$, as illustrated in
Fig.~\ref{model}a.  The volume
exclusion between rods is then modelled by a shifted and truncated
Lennard-Jones potential
\begin{equation}
V(r)=\left\{\begin{array}{l l}
4\epsilon\left[\left(\frac{l_b}{r}\right)^{12} -
       \left(\frac{l_b}{r}\right)^6\right]+\epsilon, &r<2^{1/6}l_b\\
0, &r\geq2^{1/6}l_b
\end{array}\right. \ ,
\label{eq:LJpotenttial}
\end{equation}
between beads belonging to different rods,
where $r$ is the distance between two beads, $l_b$ is the bead diameter,
and $\epsilon$ is the strength of the potential. We use $\epsilon$ as
the energy scale in our SPR simulations.

A single rod without noise then moves with a constant velocity
$v_0=F_{rod,0}/\gamma_{\parallel}$. In the non-zero noise regime, the
diffusion constant along the parallel direction is
$D_{\parallel}=\sigma_{rod}^2L_{rod}\Delta t_{rod}/2\gamma_{\parallel}^2$.
The dimensionless P\'{e}clet number, which measures the ratio
of self-propelled and diffusive motion, is thus
\begin{equation}
{\rm Pe}=\frac{L_{rod}v_0}{D_{\parallel}}=
\frac{2\gamma_{\parallel} F_{rod,0}}{\sigma_{rod}^2 \Delta t_{rod}}.
\end{equation}
We use $1/{\rm Pe}\propto \sigma_{rod}^2$ to characterize the
strength of the environmental noise.

In SPR systems \cite{Peruani2006,Baskaran2008}, alignment is
naturally introduced by the volume exclusion between the anisotropic
particles; this also implies that the interaction neighborhood needs
no further assumptions, but is directly related to the rod length.
Hard-core interactions do not distinguish the two ends of an
symmetrically elongated object. Thus, both parallel and
anti-parallel velocity configurations are induced. In simulations
of point-like SPPs, noise is implemented by adding a random
component to the velocity {\em orientation} of each particle. In our
model of SPR, random forces are applied on each rod, which results
in fluctuations in both the magnitude and the orientation of the
velocity vectors. For a single rod, the orientation fluctuations
lead to rotational diffusion, which implies a persistence length
\begin{equation}
L_p = \frac{2v_0\gamma_{\parallel}^2}{\sigma_{rod}^2L_{rod}\Delta t}
\end{equation}
of its trajectory.
Note that the noise forces are not caused by thermal fluctuations,
which would require a factor two between the variance of the random forces in
parallel and perpendicular directions. In most biological and synthetic
rSPP systems, thermal fluctuations are indeed negligible due to large size
of the particles. In these systems, the environmental noise arises,
for example, from density fluctuations of signalling molecules for
chemotactic swimmers, or from fluctuations of the motor activity.

We use rods of length $L_{rod}=11l_b$ and undisturbed velocity
$v_0=1.21 \epsilon / (\gamma L_{rod})$. Effects of a polydispersity
of rod lengths or a distribution of propulsion forces are not considered.
The motion of rods are calculated with a discrete time step
$\Delta t_{rod}=0.001$.
Most of our rod simulations start from random initial
states, where the rods are placed into the simulation box with random
orientations and random positions without overlap. If not
explicitly mentioned, the size of the simulation box is
$L_x=L_y=400l_b$, which is much larger than the rod length.
Periodic boundary conditions are employed.

Our model differs from the model of Ref.~\cite{Peruani2006}
by the type of repulsive interaction between the rods.
In Ref.~\cite{Peruani2006},
rods interact by a ``soft" volume exclusion, where the repulsion force is
proportional to the square of overlapping area, while in our model the
interaction is a short-range Lennard-Jones potential between discretized
beads. In the limit of a large overlap energy, the two models become
equivalent.


\subsection{Flagella}
\label{sec:model_flagella}

We consider a system of $N_{fl}$ flagella of length $L_{fl}$ in a
box of size $L_x \times L_y$. Each flagellum consists of
semi-flexible string of monomers of mass $m_{fl}$, connected by
springs (see Fig.~\ref{model}b). The shape of the flagellum is
determined by the elastic energy
\begin{equation}
E = \sum_{i}\frac{k}{2l_0^2} (|{\bf R}_i| - l_0)^2 +\sum_{i}\frac
{\kappa}{2l_0^3}
      \biggl \{ {\bf R}_{i+1}-\mathfrak R(l_0c) {\bf R}_i\biggr \}^2
      + V \ .
\label{eq:fullpotential}
\end{equation}
Here, the first term is the harmonic potential generated by springs
with spring constant $k$ and rest length $l_0$. ${\bf R}_i$ is the bond
vector pointing from monomer $i$ to monomer $(i+1)$. The second
term of Eq.~(\ref{eq:fullpotential}) is the bending energy the flagellum,
with bending rigidity $\kappa$. $\mathfrak R(l_0c)$ is an operator
which rotates a two-dimensional vector clockwise by an angle $l_0c$.
The local spontaneous curvature $c$ varies with time $t$ and
position $x$ along the flagellum to generate a propagating bending
wave,
\begin{equation}
    c(x,t)=A\sin\biggl[-2\pi f t+qx+\varphi\biggr] \ .
\label{eq:localcurvature}
\end{equation}
The detailed analysis of the beating pattern of nematodes
\cite{Gray1964} and bull sperm \cite{Gray1957, Riedel-Kruse2007} has
shown that a single sine mode represents the beating pattern to a
good approximation. We use the wave number $q=2\pi/L_{fl}$, such
that the phase difference between the first and the last monomer is
$2\pi$ and one complete wavelength is present on the flagellum. The
beating frequency $f$ is constant for each flagellum; it is chosen
from a Gaussian distribution, centered at $f_0$ and with variance
$\sigma_{fl}^2f_0^2$. $\varphi$ is the initial phase of the first
monomer, which is chosen from a uniform distribution in $[0,2\pi]$.
As $t$ increases, a wave propagates along the flagellum from the
first to the last monomer, pushing the fluid backwards and
propelling the flagellum forward. Although the spontaneous local
curvature $c$ is prescribed by Eq.~(\ref{eq:localcurvature}), the
flagellum is elastic and its configuration is affected by the
viscosity of the medium and the flow field generated by other
flagella. The third term in Eq.~(\ref{eq:fullpotential}) describes the
interaction between flagella due to volume exclusion; here, we
employ again the shifted and truncated Lennard-Jones potential
(Eq.~(\ref{eq:LJpotenttial})) between monomers of different flagella.

Our model of a flagellum differs from the model of a sperm employed
in Ref.~\cite{Yang2008} by the absence of a passive midpiece and a
circular head. Also, in the sperm simulations \cite{Yang2008}, two
sine waves were present on the tail, while a single sine wave is
present on the flagellum.

We use flagella of length $L_{fl}=50l_0$. The elastic moduli in
Eq.~(\ref{eq:fullpotential}) are the spring constant
$k=1.25\times10^4k_BT$ and the bending rigidity
$\kappa=200k_BTL_{fl}$. The amplitude $A=5/L_{fl}$ of the
spontaneous curvature in Eq.~(\ref{eq:localcurvature}) induces a
beating amplitude of about $6.1\, l_0=0.13\, L_{fl}$.
The strength $\epsilon=15k_BT$ of the volume exclusion is large
compared to the thermal energy. The simulations are initialized by
placing $N_{fl}$ flagella in the simulation box with random initial
positions and orientations, without any overlap. The size of the
simulation box is $L_x\times L_y$, where $L_x=L_y=400l_0$, eight
times the length of a flagellum. Periodic boundary conditions are
employed.

Each simulation run of the flagella systems covers a total time
interval of about $3300$ beats. The first $800$ beats are not taken
into account in the calculation of averages, in order to allow the
system to reach the stationary state. This time is longer than the
largest relaxation time of about $650$ beats observed in the system with
a width $\sigma_{fl}=0.1\%$ of the frequency distribution.

\subsection{Multi-Particle-Collision Dynamics (MPC)}
\label{sec:MPC}

MPC is a particle-based mesoscopic simulation technique used to
describe the hydrodynamics and flow behavior of complex fluids.
The fluid is modeled by $N_{sol}$ point particles of mass $m_{sol,i}$,
which are characterized by their
continuous space position ${\bf r}_{sol,i}$ and velocity ${\bf
v}_{sol,i}$. During every time step $\Delta t_{MPC}$, there are two
distinct simulation steps, streaming and collision. In the
streaming step, the fluid particles do not interact with each other and
move ballistically according to their velocities,
\begin{equation}
{\bf r}_{sol,i}(t+\Delta t_{MPC})=
      {\bf r}_{sol,i}(t)+{\bf v}_{sol,i}(t)\Delta t_{MPC} \ .
\end{equation}
In the collision step, the particles are sorted into the cells of
a square lattice of side length $a$ according to their position, and
interact with all other particles in same collision box through a
multi-body collision.  The collision step is defined by a rotation of all
particle velocities in a box in a co-moving frame with its center of
mass. Thus, the velocity of the $i$-th particle in the $j$-th box
after collision is
\begin{equation}
{\bf v}_{sol,i}(t+\Delta t_{MPC})={\bf v}_{cm,j}(t)+
           \mathfrak R_j(\alpha)[{\bf v}_{sol,i}-{\bf v}_{cm,j}]
\end{equation}
where
\begin{equation}
{\bf v}_{cm,j}(t)=\frac{\sum_{j} m_{sol,i}{\bf v}_{sol,i}}{\sum_{j} m_{sol,i}}
\end{equation}
is the center-of-mass velocity of $j$-th box, and $\mathfrak
R_j(\alpha)$ is a rotation matrix which rotates a vector by an angle
$\pm\alpha$, with the sign chosen at random. This implies that
during the collisions particles exchange momentum,
but the total momentum and kinetic energy
are conserved within each collision box. In order to ensure
Galilean invariance, a random shift of the collision grid has to be
performed \cite{Ihle2001}.

The total kinematic viscosity $\nu$ is the sum of two contributions,
the kinetic viscosity $\nu_{kin}$ and the collision viscosity
$\nu_\textit{coll}$. In two dimension, approximate analytical
expressions are \cite{Kikuchi2003, Ihle2003b},
\begin{eqnarray}
  \frac{\nu_{coll}}{\sqrt{k_{B}Ta^2/m_{sol}}} &=&
    \frac{1}{12 h} \, {(1-\cos{\alpha})}\, \biggl(1-\frac{1}{\rho}\biggr) \\
  \frac{\nu_{kin}}{\sqrt{k_{B}Ta^2/m_{sol}}} &=&
    h \, \biggl[\frac{1}{1-\cos\alpha}\frac{\rho}{\rho-1}-\frac{1}{2}\biggr]
\end{eqnarray}
where $\rho$ is the average particle number in each box and
$h=\Delta t_{MPC}\sqrt{k_BT/m_{sol}a^2}$ is the rescaled mean free
path. We use $k_BT=1$, $m=1$, $a=1$, $\Delta t_{MPC}=0.025$,
$\alpha=\pi/2$, and $\rho=10$. This implies, in particular, that the
simulation time unit $(ma^2/k_BT)^{1/2}$ equals unity. With these
parameters, the total kinematic viscosity of fluid is
$\nu=\nu_{coll}+\nu_{kin}\approx3.02$.

During the MPC streaming step, the equations of motion of the
flagella monomers are integrated using a velocity-Verlet algorithm,
with a molecular-dynamics time step
$\Delta t_{fl}=\Delta t_{MPC}/50 = 5\times10^{-4}$.
The bond length between the monomers is related to the collision cell
size by $l_0=a/2$.
The flagella only interact with the fluid during
the MPC collision step. This is done by sorting the flagella
monomers together with the fluid particles into the collision cells
and rotating their velocities relative to the center-of-mass
velocity of each cell. Since energy is continuously injected into
the system by the actively beating flagella, we employ a thermostat
to keep the fluid temperature constant by rescaling all
fluid-particle velocities in a collision box relative to its
center-of-mass velocity after each collision step.

With the parameters given above, a single flagellum with
$f_0=1/120$ swims forwards with the velocity
$v_{single}=0.020\pm0.001$ in a MPC fluid. Thus, we estimate a
Reynolds number $Re=2A_{fl} v_{single}/\nu\approx 0.04$ for our
flagellum model, where $A_{fl}=0.12L_{fl}$ is the beating amplitude.
The velocity of our flagella can be compared with the velocity of an
{\em infinitely long} string beating in a two-dimensional fluid at
$Re=0$, which was calculated analytically by Taylor
\cite{Taylor1951} to be
\begin{equation}
v_{single} = \frac{2\pi A_{fl}^2v_{wave}}{\lambda_{wave}^2}
   \left(1-\frac{19}{4}\frac{\pi^2 A_{fl}^2}{\lambda_{wave}^2}\right)
\end{equation}
where $\lambda_{wave}$ is the wave length and $v_{wave}=\lambda_{wave}f$
is the propagation velocity of the sinusoidal wave on the flagellum.
Applying the parameters in our simulations, we obtain $v_{single} = 0.0183$,
in excellent agreement with the simulation result. This demonstrates
that the simulation model describes the limit of low-Reynolds-number
hydrodynamics very well.

%
%
\section{Swarming Behavior of Self-Propelled Rods}
\label{sec:swarming_rods}

After starting from a random initial state, the rods
aggregate and form clusters. Large clusters can form by
collisions of smaller ones, while at the same time they can break
up due to collisions with other clusters or due to the noisy
environment.  After a transient
phase, the system reaches a stationary state, in which the
formation rate of any cluster size equals its break-up rate.
The degree of aggregation in the system depends on
its parameters such as the P{\'e}clet number and the number density
$\rho_{rod}=N_{rod}/L_xL_y$.

We define a cluster as follows. We consider two rods to be in the
same cluster if the angle between their orientation vectors is
less than $\pi/6$ and the
nearest distance is less than $2l_b$, which is about two times the
width of a rod.  A cluster is defined as a set of rods that
are neighbors either directly or through other rods at
a given moment in time. Its size is simply the number of rods it
contains. A freely gliding rod without any neighbor is considered as
a cluster of size $n=1$.

We study systems at intermediate densities, where $\rho_{rod}$ is
neither very low, such that there are hardly any collisions, nor high
enough for a nematic phase to appear for rods in
thermal equilibrium, {\em i.e.} densities lower than the critical
density $\rho_c=3\pi/(2L_{rod}^2)$ of the isotropic-to-nematic
phase transition \cite{Kayser1978}.

The statistical quantities, which will be analyze
in Secs.~\ref{sec:swarming_rods} and \ref{sec:swarming_flagella},
are listed in Table~\ref{table1}.

\begin{table}
\begin{ruledtabular}
\begin{tabular}{llll}
                    & power law        &  rods             & flagella          \\
$\Pi(n)$            & $n^\beta$        &  $-6<\beta\simeq -2$ & $-4<\beta\simeq -2$ \\
$\langle n \rangle$ & ${\rm Pe}^{\zeta}$ &  $\zeta=0.37$      &         \\
$\langle n \rangle$ & $\sigma_{fl}^{-\zeta}$ &             & $\zeta=0.26$        \\
$T_{life}$          & $n^{-\delta}$    &  $\delta=0.2$     & $\delta=0.5$
\end{tabular}
\end{ruledtabular}
\caption{Definition of power-law exponents for the cluster size
distribution $\Pi(n)$, the average cluster size $\langle n \rangle$,
and the cluster lifetime $T_{life}$, and their typical values for
rods and flagella.}
\label{table1}
\end{table}

\subsection{Cluster-Size Probability Density Functions and Stationary States}
\label{sec:PDF_rods}

\begin{figure}
\subfigure{\resizebox{4cm}{!}{\includegraphics{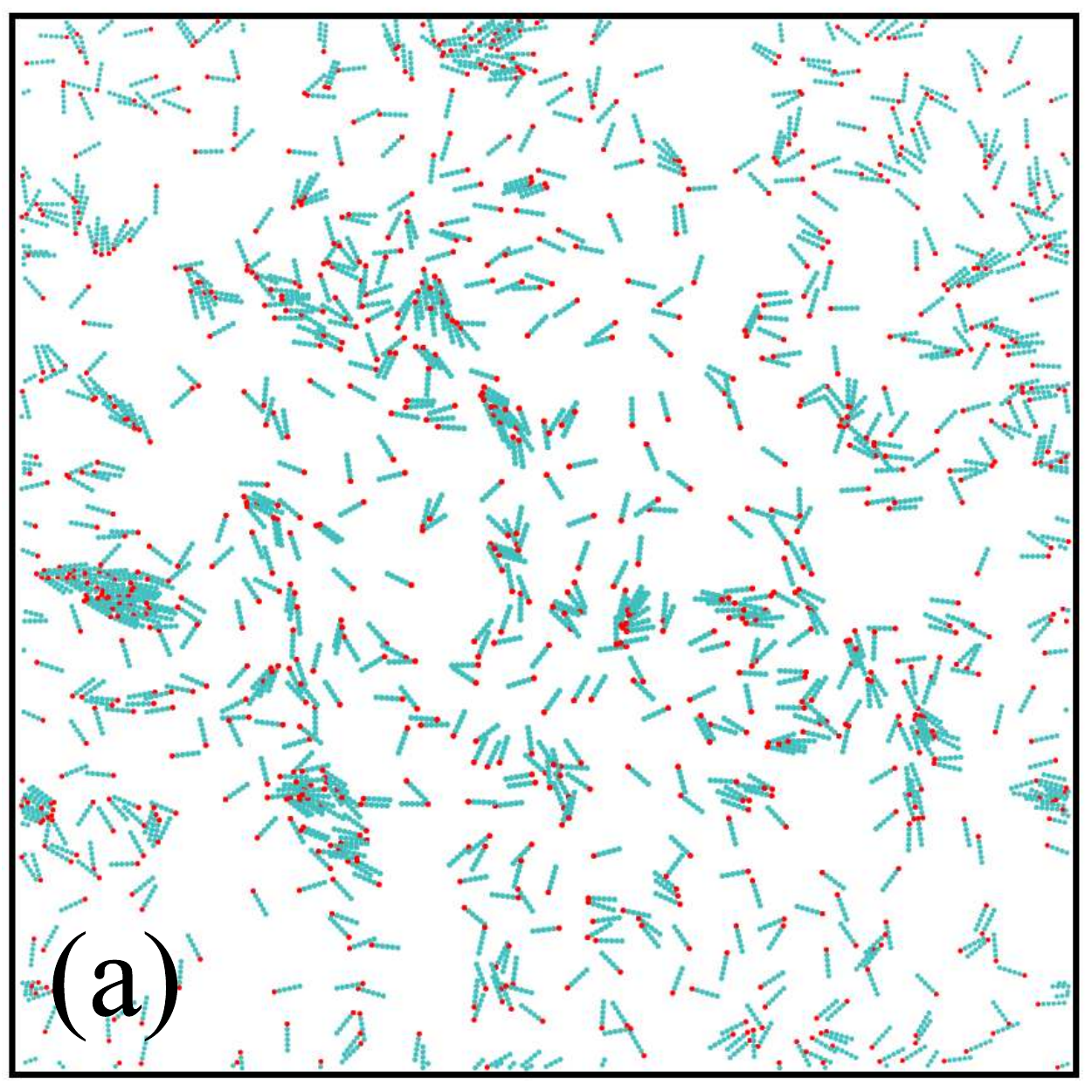}}}
\subfigure{\resizebox{4cm}{!}{\includegraphics{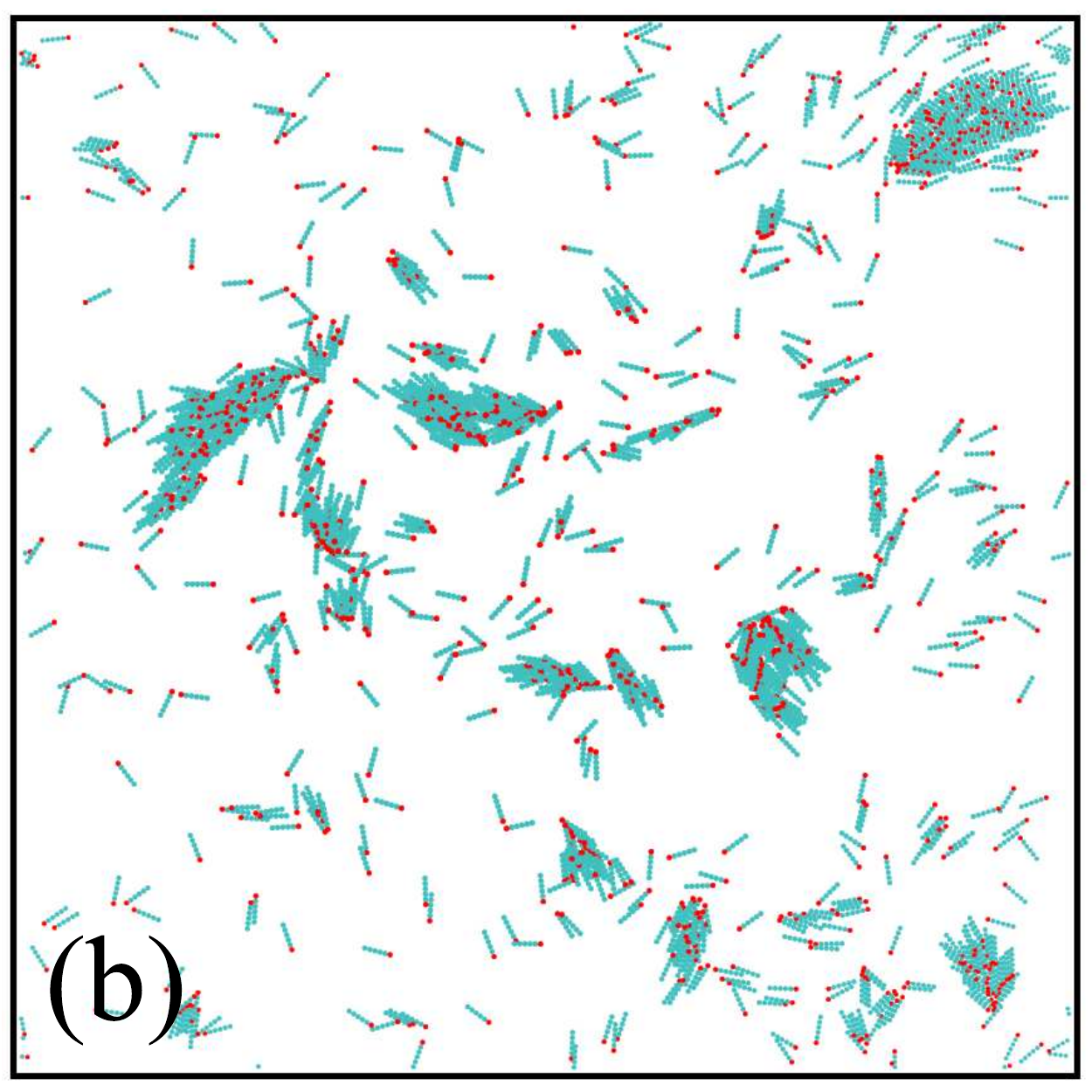}}}
\subfigure{\resizebox{4cm}{!}{\includegraphics{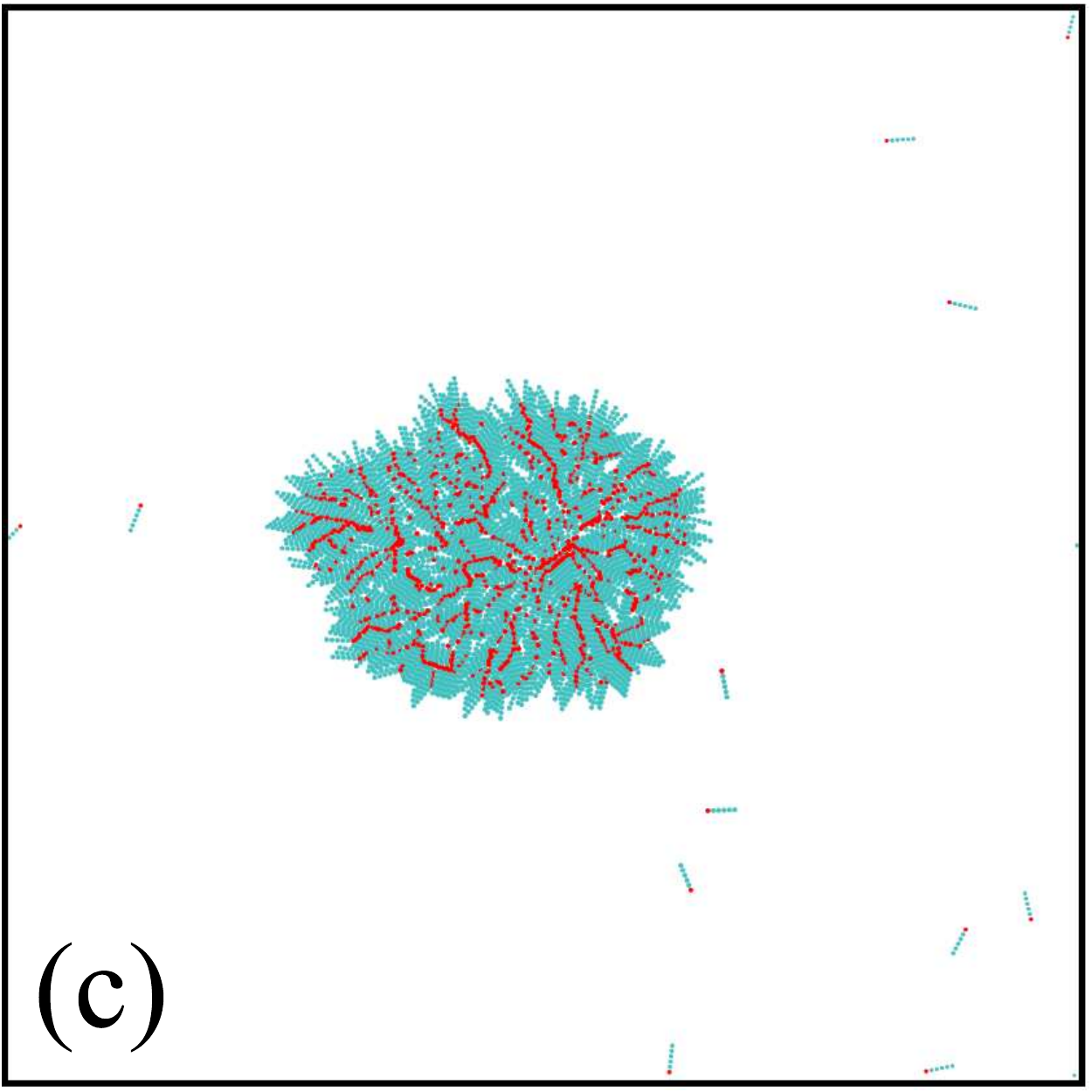}}}
\subfigure{\resizebox{4cm}{!}{\includegraphics{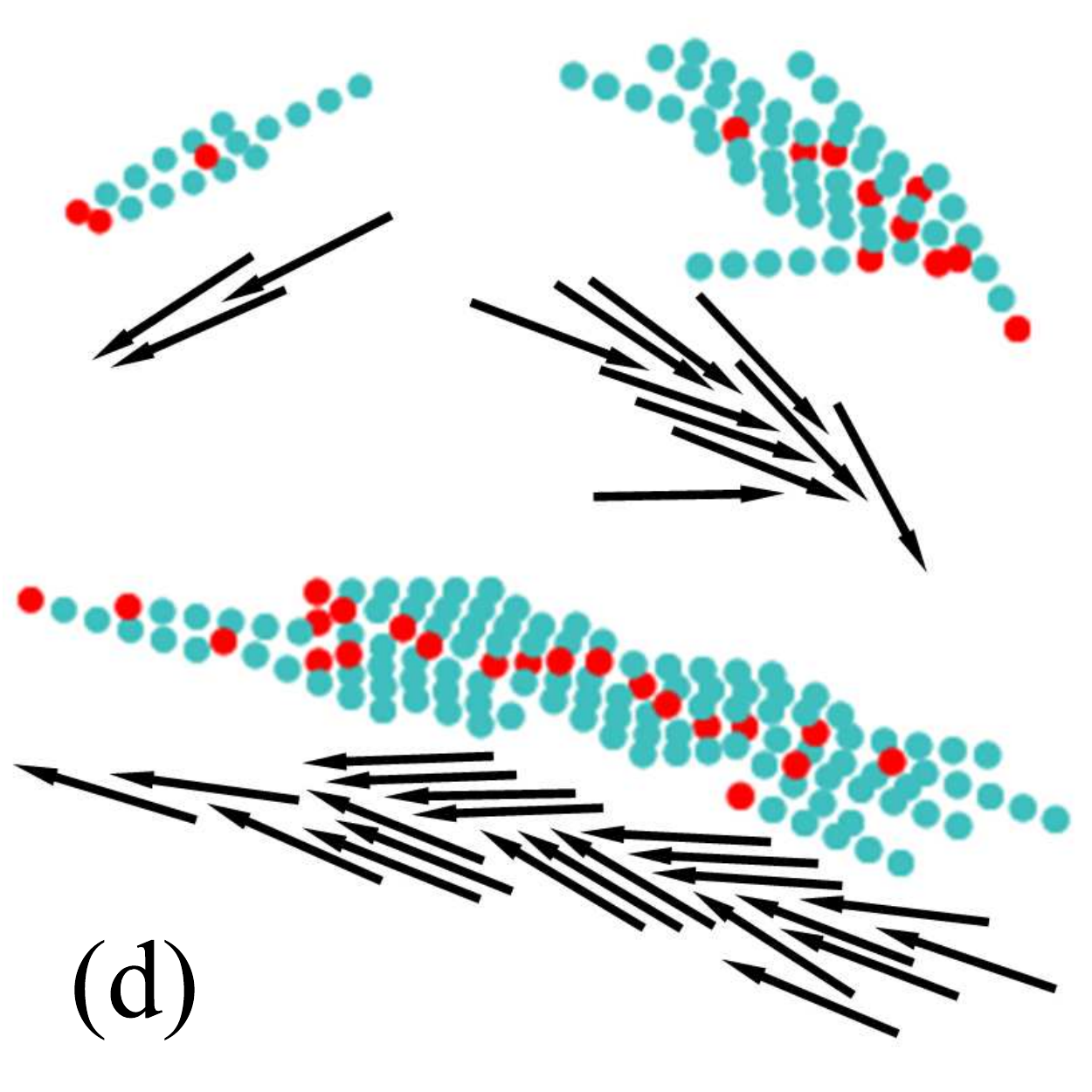}}}
\caption{(Color) Snapshots of the SPR systems at different
stationary states. Parameters are $\rho_{rod}L_{rod}^2=0.7744$ and
(a) $1/{\rm Pe}=0.02645$ ($\Pi_1$); (b) $1/{\rm Pe}=0.00501$
($\Pi_2$); (c) $1/{\rm Pe}=0.00041$ ($\Pi_3$). Red dots mark the
front ends of the rods. (d) Close-up of clusters of size $n=3$, $10$
and $22$ shows the partially blocked structure; chosen from a
simulation with parameters $\rho_{rod}L_{rod}^2=0.7744$ and $1/{\rm
Pe}=0.00095$. For a movie see Ref.~\protect\cite{movie}. }
\label{snapshot}
\end{figure}

For a system with particles distributed at random in space, the probability
of finding $n$ particles in some area obeys a binomial distribution; in our
SPR systems, the probability to find large particle numbers $n$
is increased by aggregation and clustering. The
stationary cluster-size probability density function (PDF) $\Pi(n)$ results
from the balance between the cluster formation and break-up rates.
While the former depends on the collision rate of clusters, the
latter depends also on the environmental noise and the cluster size.
We distinguish three different
stationary states in our SPR systems by comparing the shapes of
their corresponding PDFs. Snapshots are shown in
Fig.~\ref{snapshot}, a movie can be found in Ref.~\cite{movie}.

\begin{figure}
\subfigure{\resizebox{7.3cm}{!}{\includegraphics{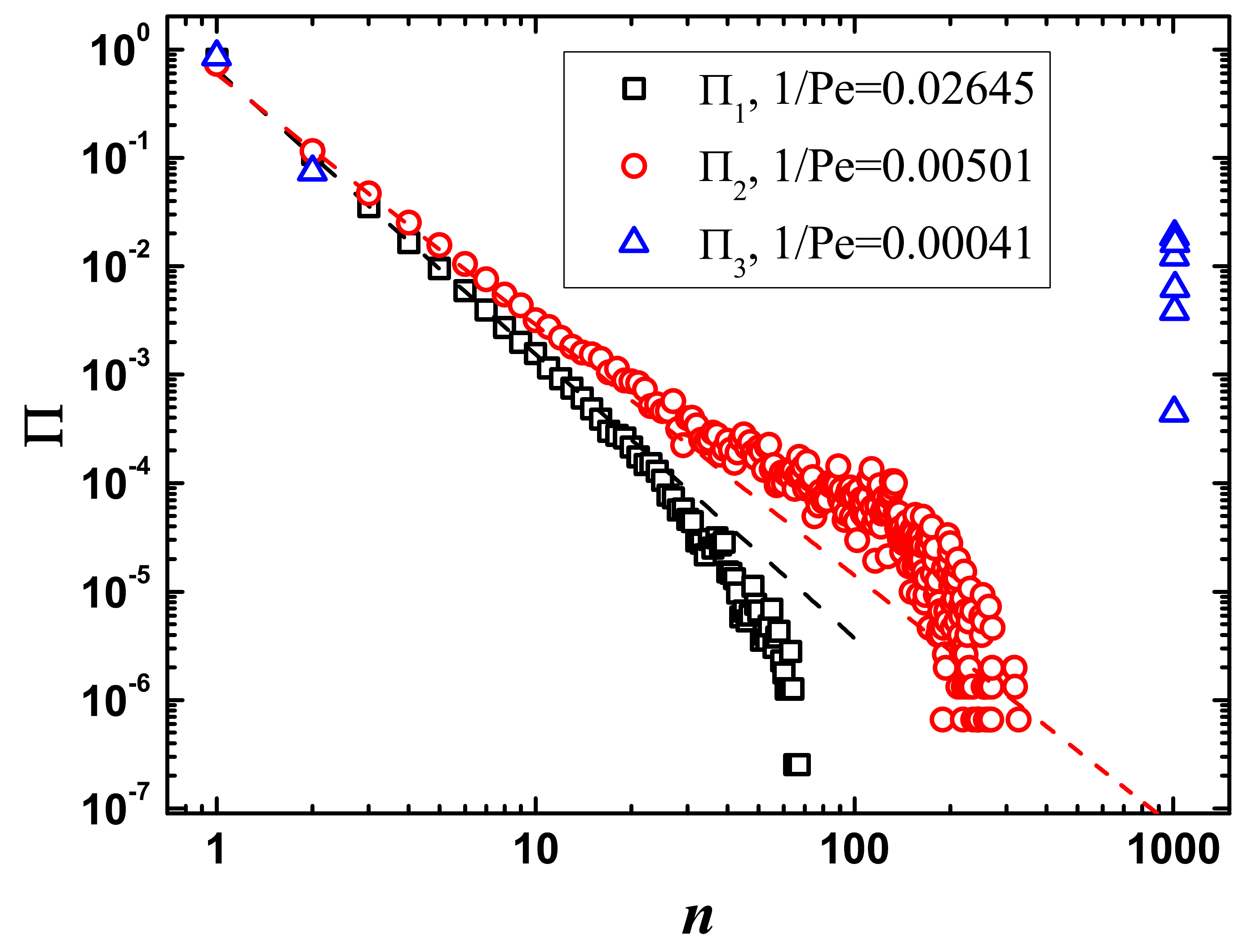}}}
\caption{(Color online) Cluster-size distribution functions $\Pi(n)$
for systems shown in the snapshots of Fig.~\ref{snapshot}(a),
$\Pi_1$ ($\square$, black), Fig.~\ref{snapshot}(b), $\Pi_2$
($\bigcirc$, red), and Fig.~\ref{snapshot}(c), $\Pi_3$ ($\triangle$,
blue). } \label{fig:PDF_rods}
\end{figure}

A disordered state, where rods are distributed in the whole
space and oriented in different directions, is characterized by a
PDF denoted as $\Pi_1$ in Fig.~\ref{fig:PDF_rods}. In a snapshot
(Fig.~\ref{snapshot}a), a weak aggregation tendency can be recognized
in this case, where several small clusters
of well polarized members glide in arbitrary directions. $\Pi_1$
decreases as a power law for small cluster sizes, then decays
exponentially for large $n$. The same kind of PDF has also been
found in simulations of swarms of point-like SPP interacting via a
phenomenological alignment mechanism \cite{Huepe2004,Huepe2008}.
The range of the power-law-decay regime of $\Pi_1$ depends on the rod
density and the environmental noise. Increasing density or
decreasing noise shifts the exponential cut-off to larger $n$.

The system with the second type of PDF, denoted $\Pi_2$ in
Fig.~\ref{fig:PDF_rods}, is more ordered, with an obvious
tendency to form large clusters. A snapshot (Fig.~\ref{snapshot}b)
shows several large and motile clusters
moving in different directions. $\Pi_2$ also displays a
power-law decay at small cluster sizes, but shows
an increased probability (compared to the power-law decay) of
finding large clusters. Increasing the number density or decreasing
the noise shifts the prominent shoulder to larger cluster sizes. For
very large aggregates, greater than the shoulder location, $\Pi_2$
decreases rapidly.

The system with the third type of PDF, denoted $\Pi_3$ in
Fig.~\ref{fig:PDF_rods}, is characterized by a giant cluster, in which
most rods are oriented radially towards the center
(Fig.~\ref{snapshot}c).
The giant cluster forms when several smaller motile clusters collide
head-on in a short time interval, such that a nucleus with a blocked
structure emerges.  This nucleus continues to grow until most of
rods in the system are gathered in it.
$\Pi_3$ has two parts, a peak at large $n$ representing the giant
clusters, and another peak at very small $n$ corresponding to some
freely swimming rods not collected by the giant cluster.
The average rod density outside the giant clusters is very low.

\begin{figure}
\subfigure{\resizebox{7.5cm}{!}{\includegraphics{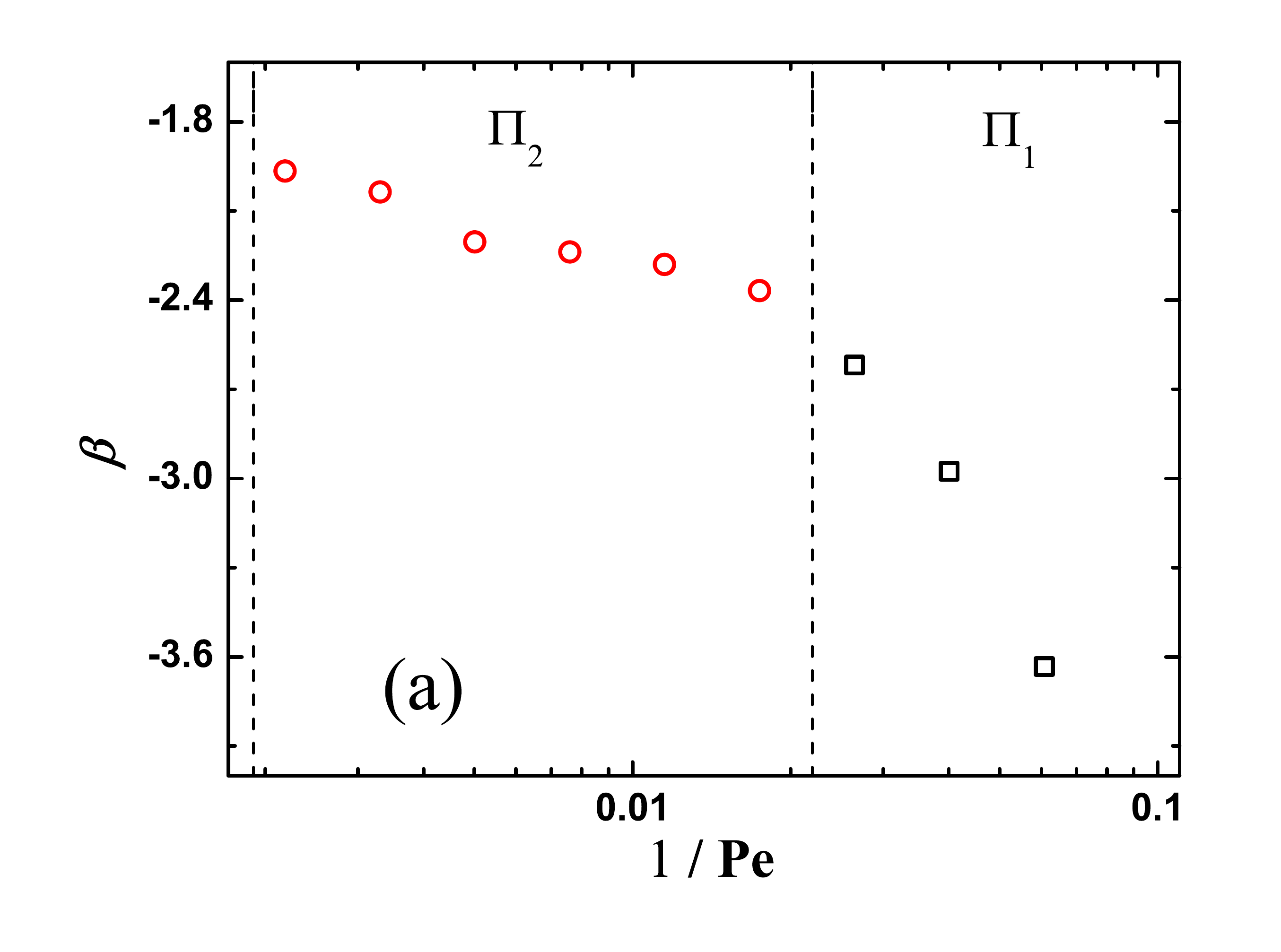}}}
\subfigure{\resizebox{7.2cm}{!}{\includegraphics{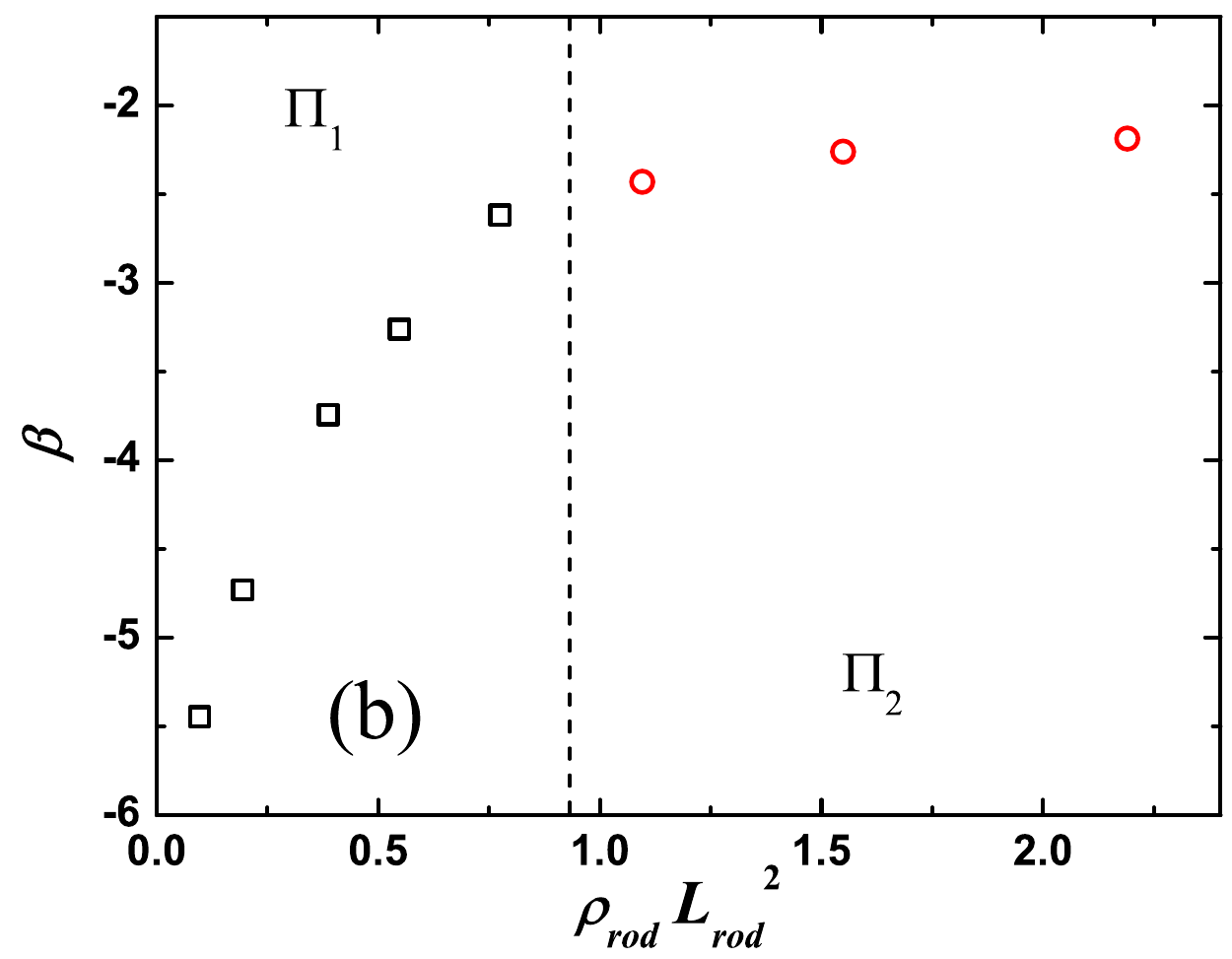}}}
\caption{The exponent $\beta$ of the power-law part of $\Pi_1$ and
$\Pi_2$ as a function of (a) the environmental noise $1/{\rm Pe}$
when $\rho_{rod}L_{rod}^2=0.7744$, and (b) the rod density
$\rho_{rod}L_{rod}^2$ when $1/{\rm Pe}=0.02645$.} \label{fig:beta}
\end{figure}

Both $\Pi_1$ and $\Pi_2$ display a power-law decay at small
cluster sizes,
\begin{equation} \label{eq:PDF_powerlaw}
\Pi \sim n^{\beta} \ .
\end{equation}
The exponent $\beta$ is a function of the rod density
$\rho_{rod}$ and noise $1/{\rm Pe}$; it
increases with increasing $\rho_{rod}$ and decreases with
increasing $1/$Pe (Fig.~\ref{fig:beta}). However, the dependence of
$\beta$ on $\rho_{rod}$ or $1/$Pe in the $\Pi_1$ regime is
much stronger than in the
$\Pi_2$ regime; in the latter case, $\beta$ approaches $-2$.

\begin{figure}
\includegraphics[width=7cm]{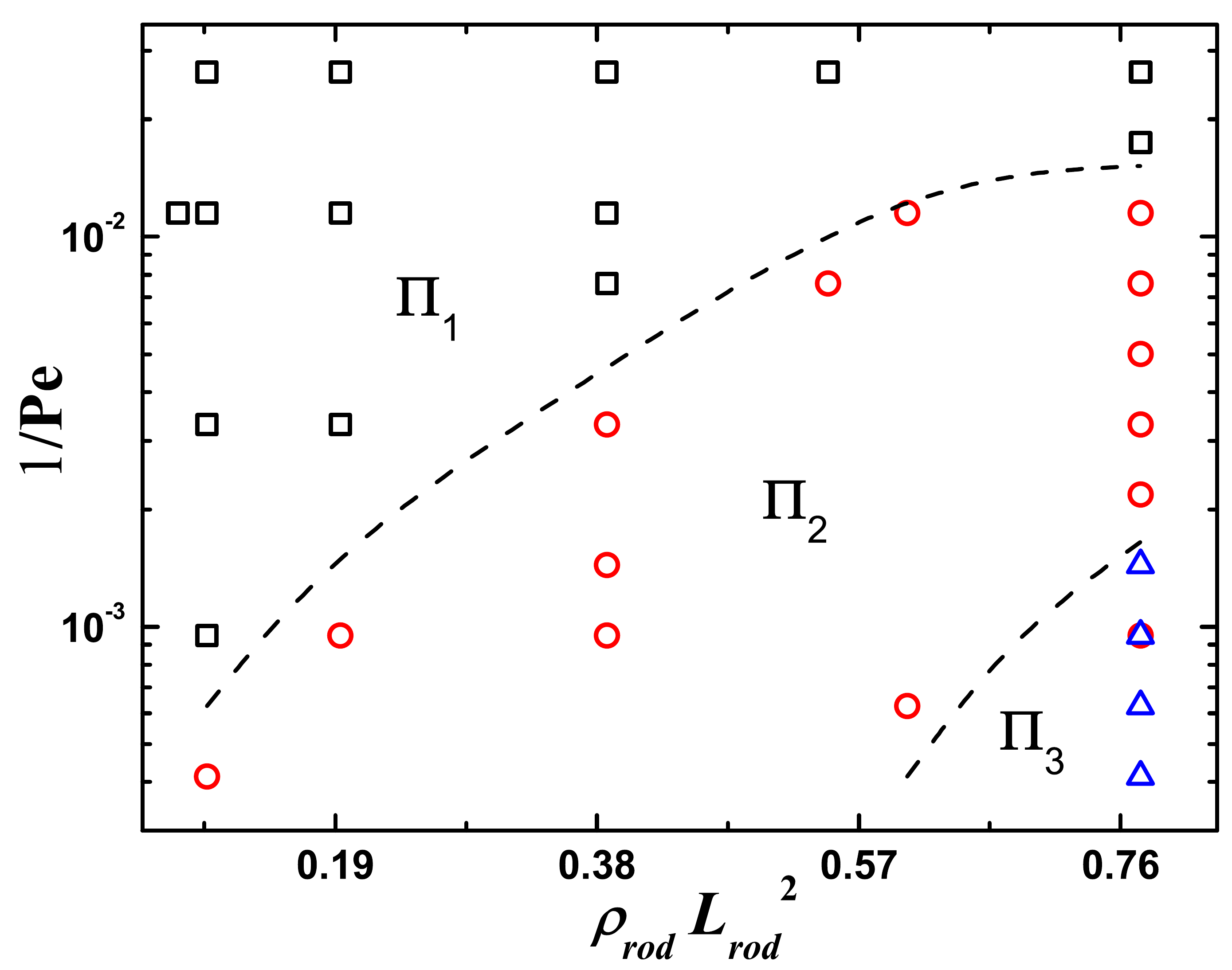}
\caption{(Color online) Dynamical phase diagram of swarm behavior.
Symbols indicate
systems with PDF $\Pi_1$ ($\square$, black), $\Pi_2$ ($\bigcirc$, red)
and $\Pi_3$ ($\bigtriangleup$, blue).
All systems were started from a random initial state. The dashed
lines are guides to the eye.}
\label{fig:phasediagram}
\end{figure}

By systematically varying the rod density $\rho_{rod}$ and the
environmental noise level, we can construct a phase diagram with
regions characterized by different types of PDFs, see
Fig.~\ref{fig:phasediagram}. Clearly, $\Pi_1$ is found in the
low-density and high-noise regime, $\Pi_3$ in the high-density and
low-noise regime, and $\Pi_2$ is associated with the transition region
between $\Pi_1$ and $\Pi_3$. Note that all systems in
Fig.~\ref{fig:phasediagram} were started from disordered initial states.
Systems characterized by the probability density function $\Pi_2$
bear some similarity with
liquid systems supercooled below the freezing point. Note the system
with $1/{\rm Pe}=0.00095$ and $\rho_{rod} L_{rod}^2=0.7744$ in
Fig.~\ref{fig:phasediagram} displays both
$\Pi_2$ and $\Pi_3$ distributions corresponding to simulations
with different initial random states. Systems with the
probability density function $\Pi_3$ show the characteristics of a glassy
behavior, where the dense packing of rods arises from the random
collisions, and remains frozen at later times.

Our results are consistent with those of
Ref.~\cite{Peruani2006}. By comparing short runs for systems with
and without fluctuations, the transition from $\Pi_1$ to $\Pi_2$ was
found in Ref.~\cite{Peruani2006} to shift to larger values of the
aspect ratio $L_{rod}/l_b$ and total area fraction of rods
$\eta=\rho_{rod}L_{rod}l_b$. Fig.~\ref{fig:phasediagram} demonstrates that
in our system the transition shifts with increasing $1/$Pe to larger
$\rho_{rod} L_{rod}^2$, which is proportional to $\eta L_{rod}/l_b$.

\subsection{Orientational Correlation Functions}

Although we distinguish three swarming states in our SPR systems,
there are only two types of cluster structures. The motile clusters
in the $\Pi_1$ and $\Pi_2$ states consist of polarized rods, as shown
in Fig.~\ref{snapshot}a,b.
In contrast, the giant clusters found in the $\Pi_3$ state consist of a
large number of rods blocking each other
in their forward motion, as shown in Fig.~\ref{snapshot}c.

\begin{figure}
\subfigure{\resizebox{7cm}{!}{\includegraphics{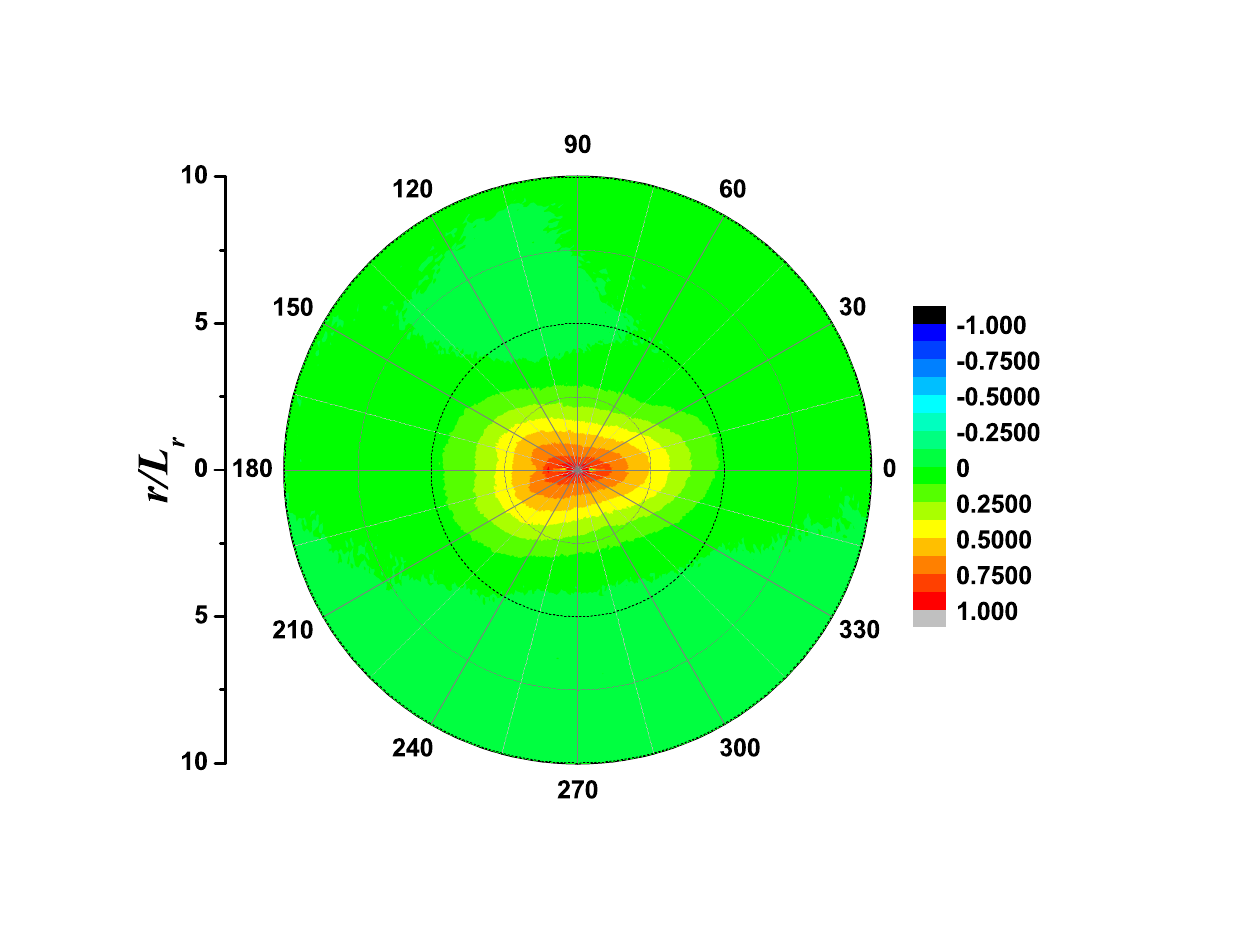}}}
\subfigure{\resizebox{7cm}{!}{\includegraphics{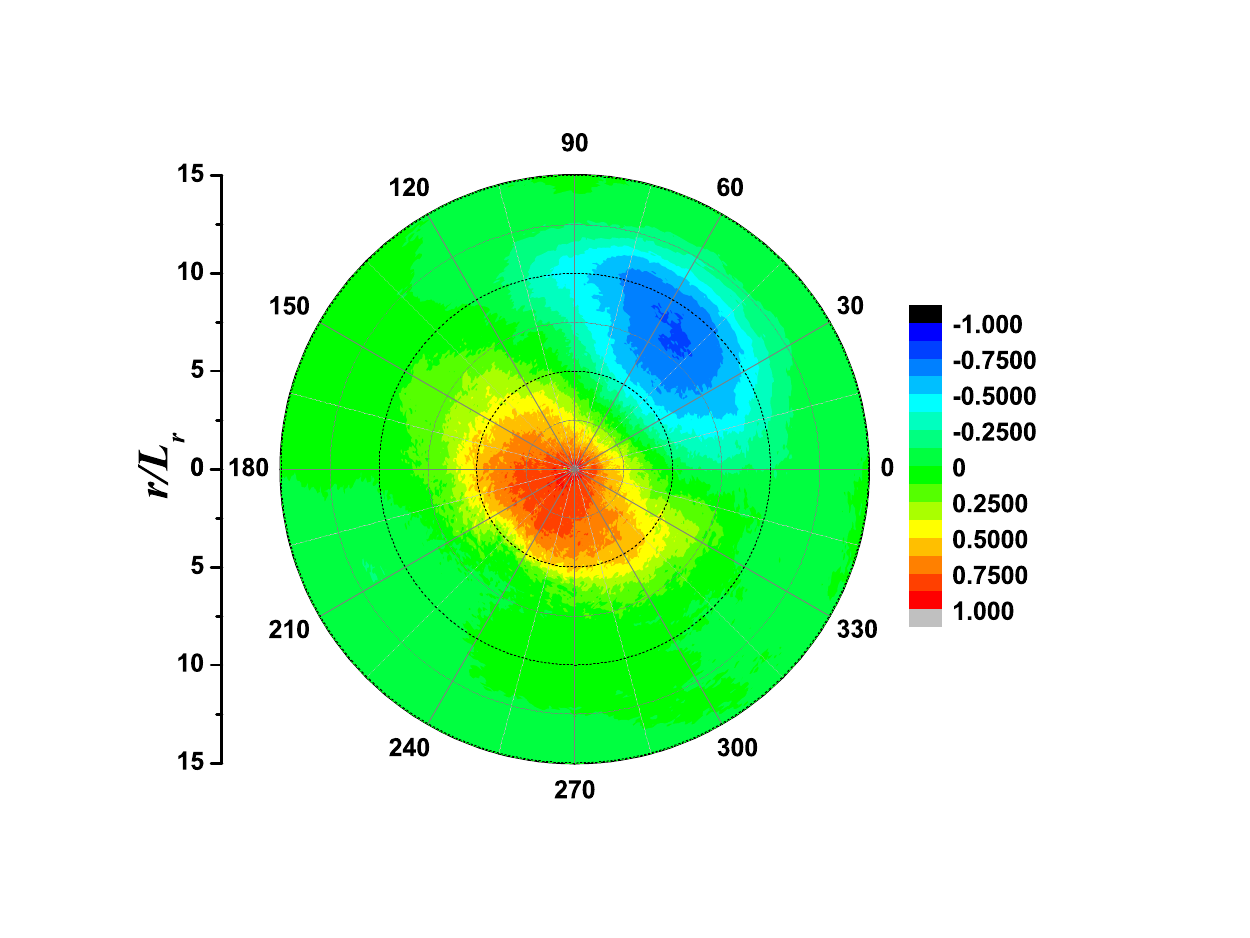}}}
\caption{(Color) The orientational correlation function $G({\bf r})$
as a function of the relative position ${\bf r}$ (a) in a system
with $\Pi_2$ ($\rho_{rod}L_{rod}^2=0.7744$ and $1/{\rm
Pe}=0.00501$), where there are motile clusters; (b) in a system with
$\Pi_3$ ($\rho_{rod}L_{rod}^2=0.7744$ and $1/{\rm Pe}=0.00041$)
characterized by the presence of a giant cluster.}
\label{correlation}
\end{figure}

These two types of clusters can be distinguished by
analyzing the orientational correlation function
\begin{equation}
G({\bf r})=\frac{1}{N_{rod}\left(N_{rod}-1\right)}
\sum_i \sum_{j\ne i} \langle{\bf \hat u}_i\cdot{\bf \hat u}_j \cdot
                     \delta({\bf r}-{\bf r}_{ij})\rangle \ .
\end{equation}
Here ${\bf\hat u}_i$ is the unit vector denoting the orientation of
rod $i$, ${\bf r}_{ij}(r,\phi)$ is the vector pointing from the
center of mass of rod $i$ to rod $j$, and $\phi$ is the angle
between ${\bf\hat u}_i$ and ${\bf r}_{ij}$. $G(r) \to 1$ for $r\to 0$
because two neighboring rods at close distance are always aligned.
At large distance, $G(r)\to 0$.

When the system is in a state characterized by $\Pi_1$ or $\Pi_2$,
$G({\bf r})$ is symmetric with respect to the direction
$\phi=0^\circ$ with a maximum at $r=0$ (Fig.~\ref{correlation}a).
The slight elongation of $G({\bf r})$ in the directions
$\phi=0^\circ$ and $\phi=180^\circ$ indicates that the clusters tend
to slightly extend in the direction of the average rod orientation
due to packaging.  The width of $G({\bf r})$ is narrower in the
front and wider in the back, because of their partially blocked
structure (see Fig.~\ref{snapshot}d) and because large clusters are more
likely to collide with other clusters head-on. If an
head-to-head collision does not result in the formation a larger
cluster or a blocked structure, the front tips are sharpened due to the
``attrition" of the two clusters.

If the system is in the state with a giant cluster, $G({\bf r})$
shows a very different behavior, see Fig.~\ref{correlation}b.
$G({\bf r})$ still has a positive maximum near $r=0$, which
represents a high local orientational order. However, a region with
negative correlations, $G<0$, develops, with a minimum at some
$(r',\phi')$. Because all rods point preferentially towards the
center of cluster, the propelling forces of the rod
nearly compensate each other. Therefore, the locomotion speed of a
giant cluster is much smaller than the gliding speed of a single
rod. Moreover, the propelling forces generate a net
torque due to the deviation of the rod orientations from pointing
exactly towards the center of mass, which implies a rotational
motion of the giant cluster. $\phi'$ is related to this rotation.
For $0^\circ<\phi'<90^\circ$, the cluster rotates counterclockwise;
for $-90^\circ<\phi'<0^\circ$, it rotates clockwise;
for $\phi'=0^\circ$, there is no net torque and the giant cluster
does not rotate.

\subsection{Average Cluster Size}

\begin{figure}
\subfigure{\resizebox{7.1cm}{!}{\includegraphics{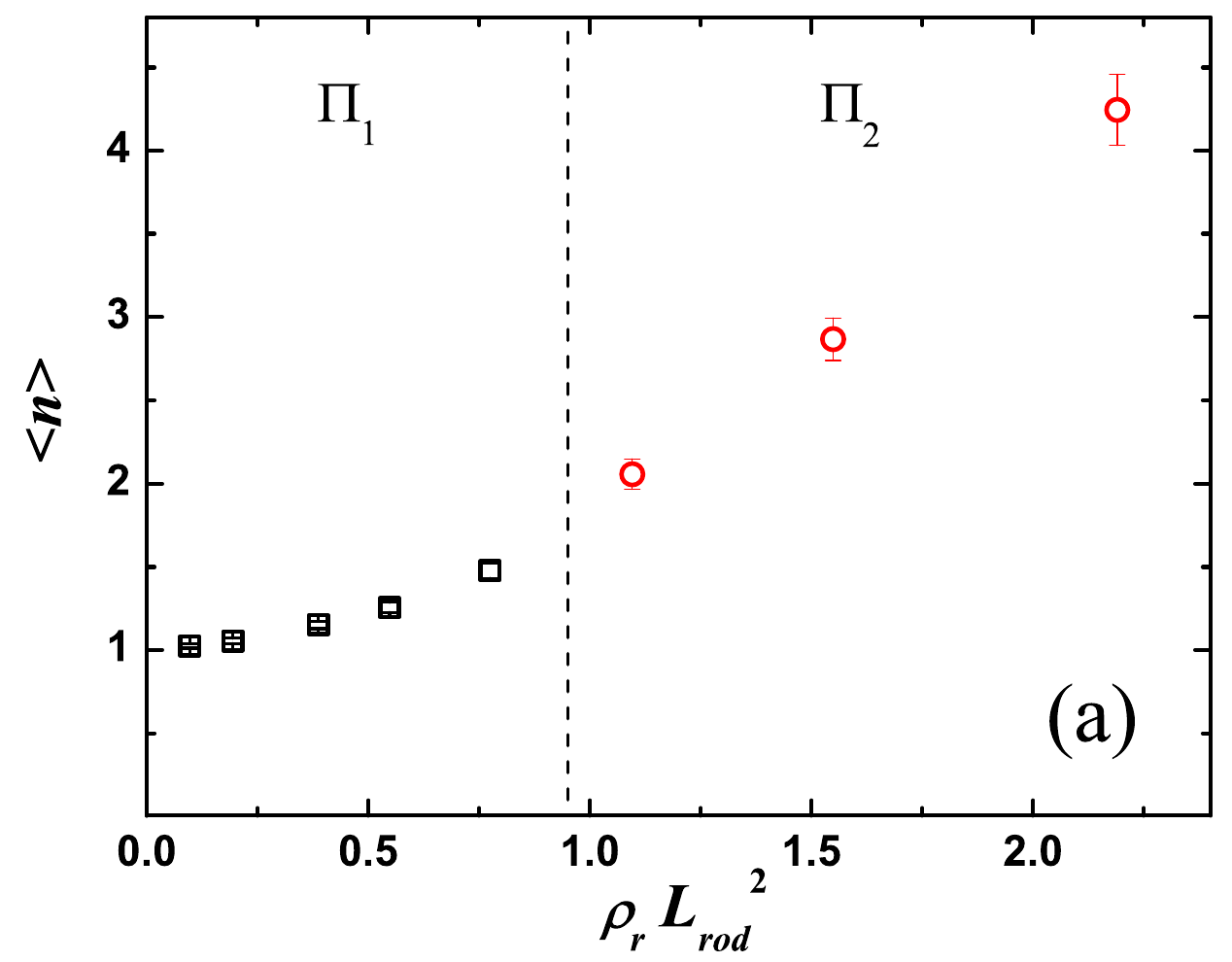}}}
\subfigure{\resizebox{7.2cm}{!}{\includegraphics{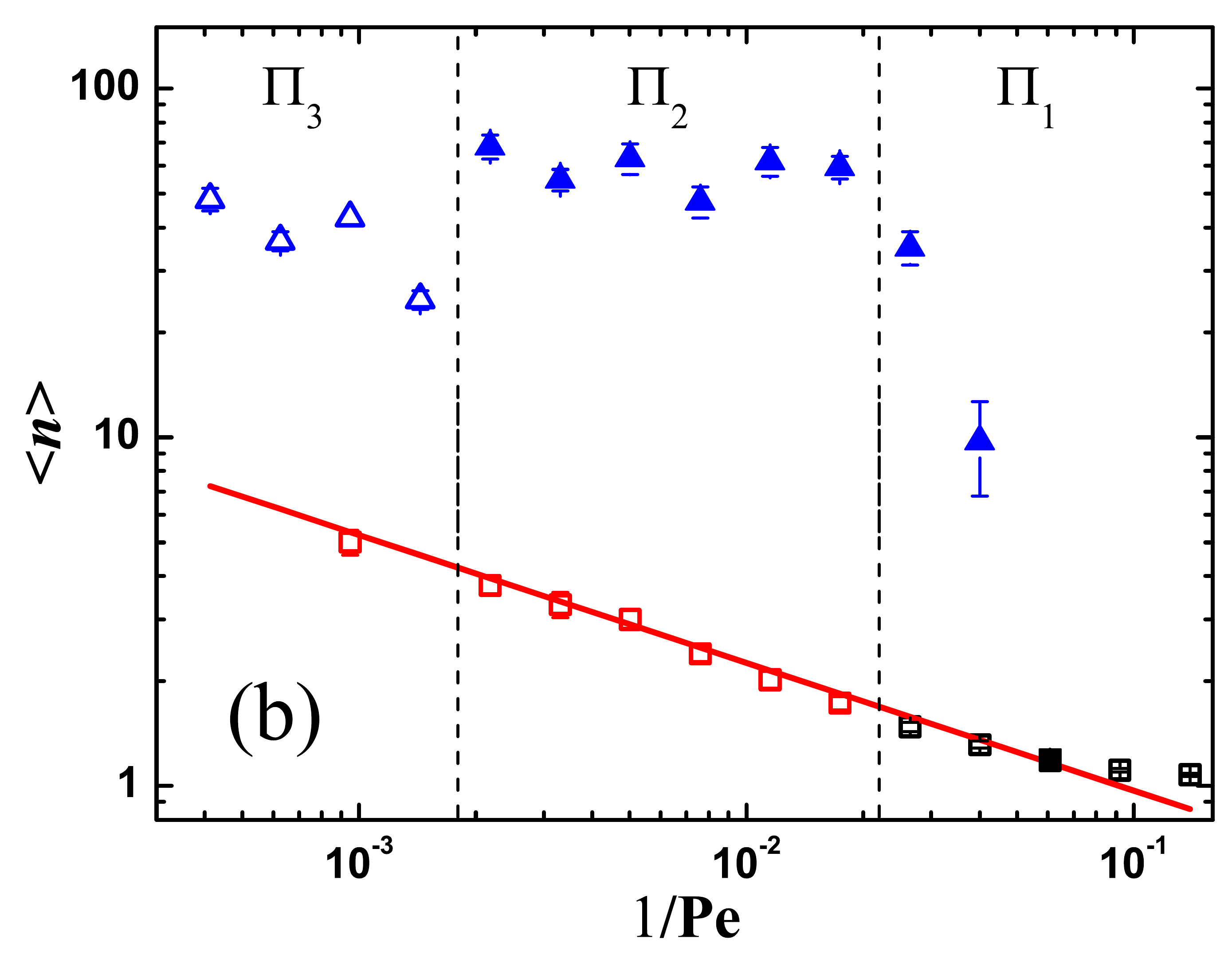}}}
\caption{(Color online) The average cluster size $\langle n\rangle$
as a function of (a) the rod density $\rho_{rod}L_{rod}^2$ with
$1/{\rm Pe}=0.02645$ and (b) the environmental noise $1/$Pe with
$\rho_{rod}L_{rod}^2=0.7744$. The solid (red) line is a fit of the
power-law part with exponent $\zeta=0.37$. The dashed (black) lines
are the boundaries separating different regions in the phase diagram
(Fig.~\ref{fig:phasediagram}). The open symbols represent systems
with $\Pi_3$ ($\bigtriangleup$, blue), $\Pi_2$ ($\bigcirc$, red),
and $\Pi_1$ ($\square$, black), starting from random initial states.
The solid symbols represent the systems with $\Pi_2$
($\blacktriangle$, blue) and $\Pi_1$ ($\blacksquare$, black),
starting from a state with a giant cluster. }
\label{averclustersize}
\end{figure}

The average cluster size $\langle n\rangle$ of the system is
\begin{equation}
\langle n\rangle=\sum_{n} n\Pi(n) \ ,
\end{equation}
where $\Pi(n)$ is the normalized cluster-size distribution function.
$\langle n\rangle$ increases with
increasing $\rho_{rod}$, as shown in Fig.~\ref{averclustersize}a; in
the low-density limit, $\langle n\rangle$ approaches unity.
$\langle n\rangle$ decreases with increasing noise level, $1/$Pe,
as shown in Fig.~\ref{averclustersize}b.
In the $\Pi_2$ regime, the system exists in two metastable states,
depending on the initial conditions.
With random initial conditions, a ``supercooled" state emerges,
which transforms into
the $\Pi_3$ state once a giant-cluster nucleus has formed.
This can be seen in Fig.~\ref{averclustersize}b for $1/{\rm Pe}=0.00095$,
where two data points
show simulation results with different random number for random
initial states.
With a giant cluster as initial state, the system stays in
the $\Pi_3$ state unless the noise is large enough to destroy the
giant cluster; this occurs in Fig.~\ref{averclustersize}b for
$1/{\rm Pe}=0.04$.
Interestingly, $\langle n\rangle$ shows a power-law decay
\begin{equation}
\langle n\rangle \sim {\rm Pe}^\zeta
\end{equation}
in the $\Pi_1$ and $\Pi_2$ region when the system starts from
a disordered state, with exponent $\zeta \simeq 0.37$.

\subsection{Cluster Lifetime}
\label{sec:lifetime_rods}

We define the lifetime of a cluster as the length of the time
during which its members do not change. The lifetime of a
cluster is analyzed with a time interval $\Delta \tau=100$; thus, cluster
lifetimes less than $\Delta \tau$ cannot be resolved.
The average cluster lifetime $T_{life}$ is a function
of cluster size $n$.

\begin{figure}
\includegraphics[width=7.1cm]{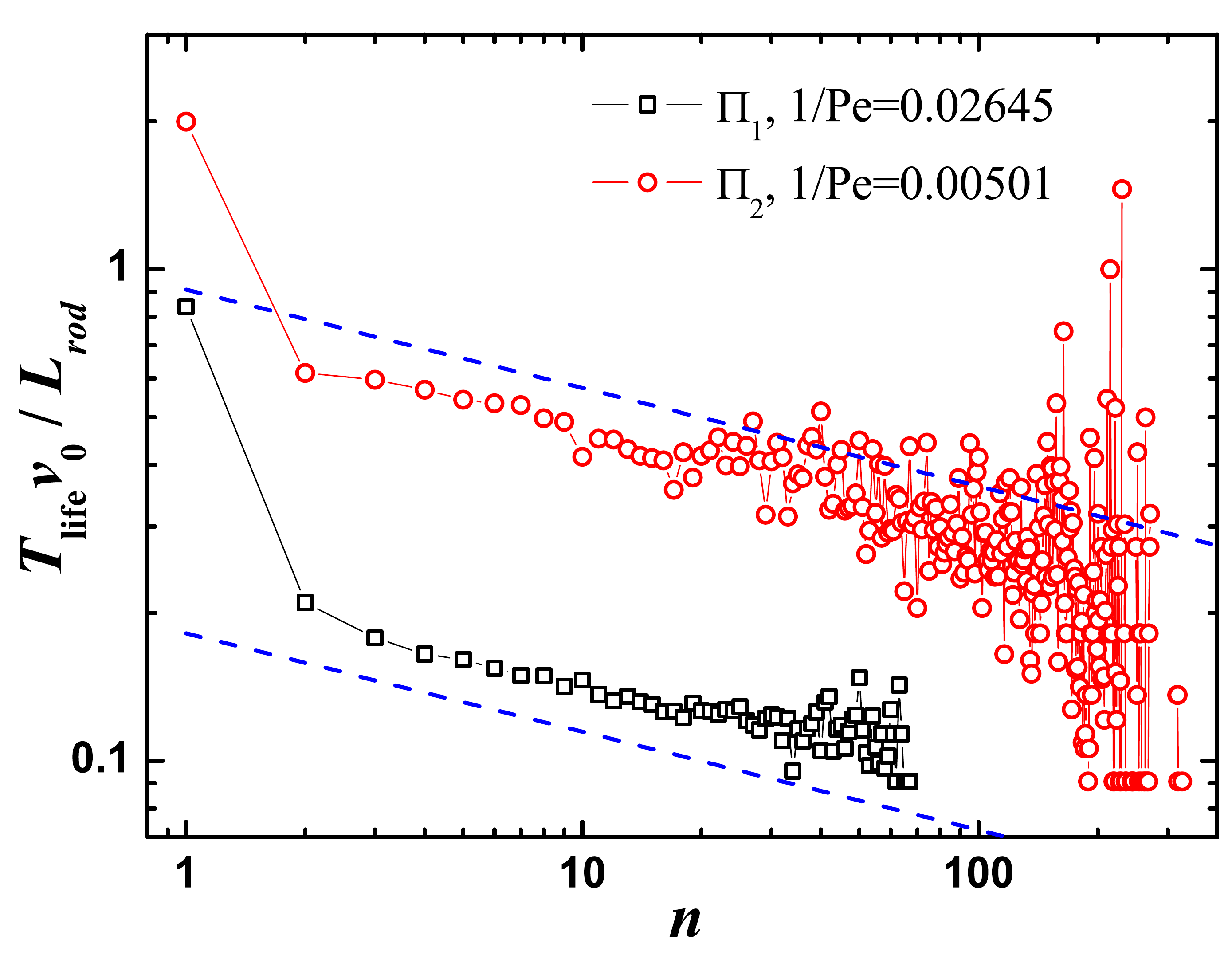}
\caption{(Color online) Average cluster lifetime
$T_{life}(n)$ for systems with the same rod number density
$\rho_{rod}L_{rod}^2=0.7744$ but with a different noise level, as
indicated. The dashed lines are power laws (\ref{eq:lifetime}) with
an exponent $\delta=0.2$.}
\label{fig:lifetimevssigma}
\end{figure}

As shown in Fig.~\ref{fig:lifetimevssigma}, the lifetimes of
the clusters of size $n=1$ are always much longer than
of other cluster sizes, because single-rods ``clusters" cannot disintegrate.
For $n\ge 2$, $T_{life}(n)$ decreases smoothly with increasing cluster size.
The data for mid-size clusters ($2<n<30$) show an
effective power-law dependence,
\begin{equation}
\label{eq:lifetime} T_{life}(n) \sim n^{-\delta}
\end{equation}
with an exponent $\delta \simeq 0.2$.
Because the environmental noise determines the break-up
rate of clusters, $T_{life}$ increases with decreasing $1/$Pe. We
only show the lifetime of motile clusters in systems characterized
by $\Pi_1$ and $\Pi_2$. The giant clusters found in the state
characterized by $\Pi_3$
can persist for a very long time until a sufficiently large
fluctuation occurs.

To understand the dependence of the cluster lifetime on $n$,
we can assume that only single rods are lost at the cluster
surface \cite{Peruani2006}.
In this case, the probability to loose
a rod per unit time is proportional to the perimeter length,
which scales as $n^{1/2}$ (for compact clusters of approximately
circular shape). Therefore, this simple argument implies a
scaling law (\ref{eq:lifetime}) with exponent $\delta=0.5$.
The growth of clusters is more complex, since
it can occur by collision with all types of other clusters;
however, the collision cross-section should again be proportional
to $n^{1/2}$.
The value of $\delta=0.5$ is considerable
smaller (corresponding to
shorter lifetimes for larger clusters) than observed in our
simulations. This indicates that there must be another
mechanism of cluster decay. Indeed, the typical cluster
configurations of Fig.~\ref{snapshot}d  indicate that only at few
places along the perimeter, rods may have the possibility to
leave the cluster.

\subsection{Finite-Size Effects}

In our simulations, the finite simulation-box size implies a finite
number of particles. A cluster can never grow larger than the total
number of rods in the system. Consequently, all quantities related
to the cluster size, such as the cluster size distribution $\Pi$ and
the stationary average cluster size $\langle n\rangle$ display
finite-size effects. Similarly, density fluctuations at the scale
of the simulation-box size are suppressed.

\begin{figure}
\subfigure{\resizebox{7.5cm}{!}{\includegraphics{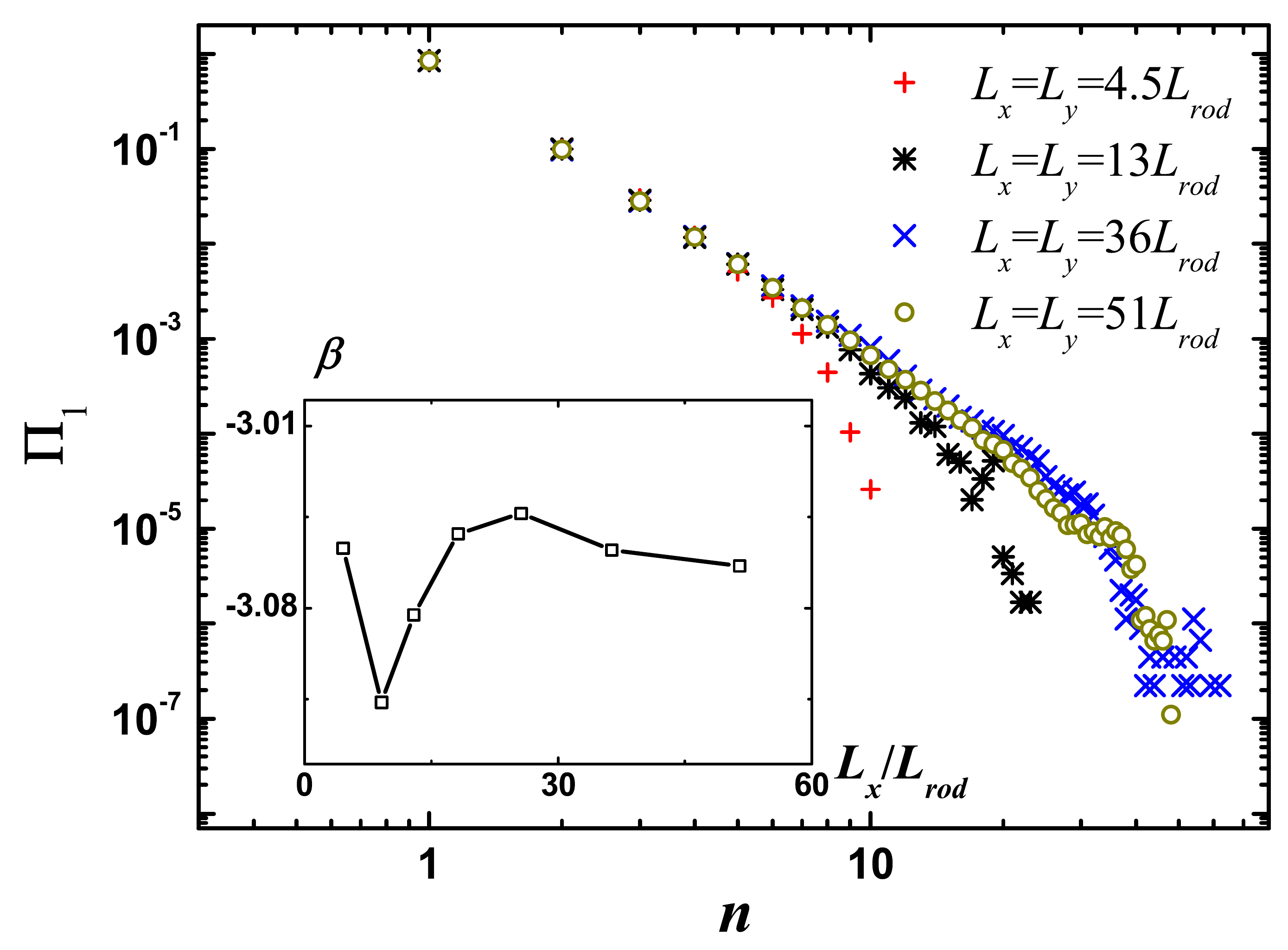}}}
\caption{(Color online) Effect of finite system size on the
probability density distribution function $\Pi_1(n)$ for
$\rho_{rod}L_{rod}^2=0.7744$, $1/{\rm Pe}=0.04009$, and different
simulation box sizes, as indicated. The inset shows the exponent
$\beta$ of the power-law part of $\Pi_1(n)$ as a function of the
size $L_x$ of the simulation box.} \label{finitePDF}
\end{figure}

For the probability density function $\Pi(n)$, the absence of
cluster larger than $N_{rod}$
does not only introduce a cut-off at large cluster size, but also
affects the exponent $\beta$ of the power-law part, as
shown in Fig.~\ref{finitePDF}.
For systems with $\Pi_1$, the data for
small box sizes ($L_x=L_y=4.5L_{rod}$ and $13L_{rod}$) still obey a
power-law decay at small $n$, without an obvious change of the
exponent, as shown in the inset of Fig.~\ref{finitePDF}, but they
deviate from the
power law when $n$ approaches $N_{rod}$. When the simulation box is
large ($L_x=L_y=36L_{rod}$ and $51L_{rod}$), the PDFs almost
coincide, and their exponential cut-offs are observed
at the same value of $n$; also, $\beta$ approaches an asymptotic value
when $L_x$ increases. Therefore, we conclude that
our results for the larger systems represent the thermodynamic limit.
Similarly, the power-law part of $\Pi_2$ extends
with increasing box size, and the location of the prominent
shoulder shifts to larger cluster size. The
finite-size effects are significantly stronger for systems in the
$\Pi_3$ region of the phase diagram. When the system is too small,
the total rod number is not sufficient to trigger the formation of
a blocked structure. The system then stays in a $\Pi_2$ state.
This supports the claim that the
state with $\Pi_2$ is a ``supercooled" state.
We believe that the absence of the $\Pi_3$ state in
Ref.~\cite{Peruani2006} is due to finite-size effect; a system of
only 100 rods is too small to form a blocked structure.

\begin{figure}
\includegraphics[width=7cm]{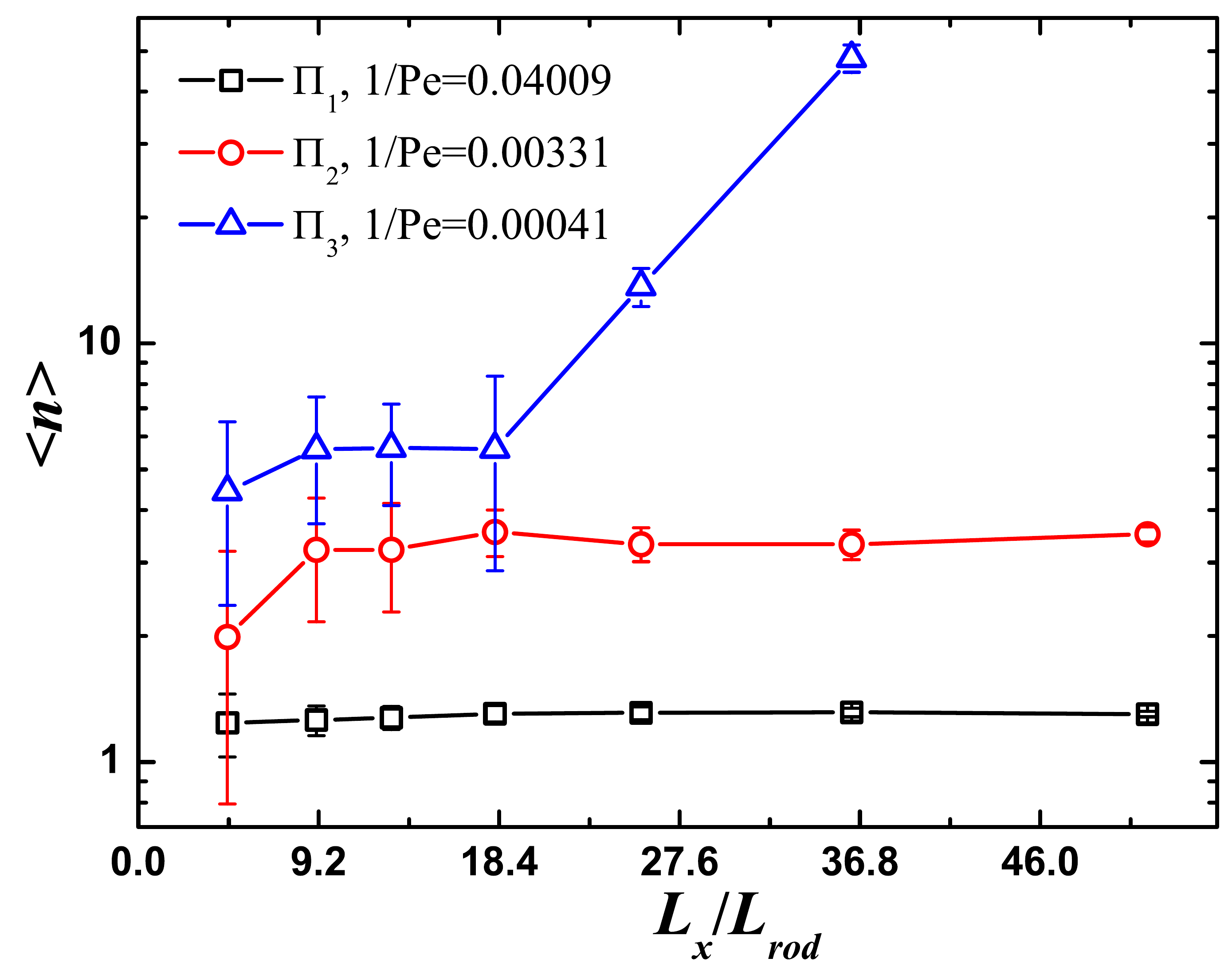}
\caption{(Color online) The average cluster size $\langle n\rangle$
as a function of the size $L_x$ of the simulation box for the
three clustering states. The number density is
$\rho_{rod}L_{rod}^2=0.7744$ in all systems.}
\label{fig:finiteavern}
\end{figure}

The dependence of the average cluster size $\langle n\rangle$ on
the linear system size $L_x$
is shown in Fig.~\ref{fig:finiteavern}. For systems with $\Pi_1$ and
$\Pi_2$, $\langle n\rangle$ increases with $L_x$ and
eventually reaches a plateau value. For the system with $\Pi_3$,
$\langle n\rangle$ strongly diverges when $L_x$ increases. Thus,
$\langle n\rangle$ can be considered as an intensive quantity in
the first two states, and as an extensive quantity in the third state.

Suppose the probability density function $\Pi(n)$ obeys a power
law for all cluster sizes,
\begin{equation} \label{eq:power_law_pi}
\Pi(n)=\frac{1+\beta}{N^{1+\beta}-1}\ n^{\beta} \ ,
\end{equation}
where $\beta<-1$ and $N=\rho_{rod}L_xL_y$ is the total number of
rods in the system. It is easy to verify that for $N\gg 1$,
where sums over $n$ can be well approximated by integrals, $\int_{1}^N
\Pi(n) dn=1$, so that $\Pi(n)$ is properly normalized. The sharp drop
due to the limited box size is neglected. In this case, the average
cluster size of the system is obtained to be
\begin{equation} \label{eq:finite_size}
\langle n\rangle=\left\{\begin{array}{l l}
-(1+\beta)N^{2+\beta}/(2+\beta), &-2<\beta<-1\\
(1-N^{-1})^{-1}\ln N, &\beta=-2\\
(1+\beta)/(2+\beta), &\beta<-2
\end{array}\right.
\end{equation}
For $-2<\beta<-1$, the average cluster size strongly depends on the
total number $N$ of rods, whereas for $\beta<-2$, $\langle n\rangle$
is independent of $N$. For large negative $\beta$, $\langle n\rangle$
approaches unity, which means that all rods are gliding freely.

In our simulations, the effective exponents in the $\Pi_1$ and
$\Pi_2$ regimes are $-6 \lesssim \beta \lesssim -2.5$ and $-2.5
\lesssim \beta \lesssim -2.0$, respectively, see
Fig.~\ref{fig:beta}. Thus, Eq.~(\ref{eq:finite_size})
implies that finite-size effects are weak in the $\Pi_1$ regime, and
are pronounced in the $\Pi_2$ regime, in agreement with the simulation
results of Fig.~\ref{fig:finiteavern}.
(Eq.~(\ref{eq:finite_size}) does not apply to the $\Pi_{3}$ state since
the assumption of a power-law dependence (\ref{eq:power_law_pi}) does
not hold.)

%
%
\section{Swarming Behavior of Flagella in a MPC Fluid}
\label{sec:swarming_flagella}

\begin{figure}
\includegraphics[width=5cm]{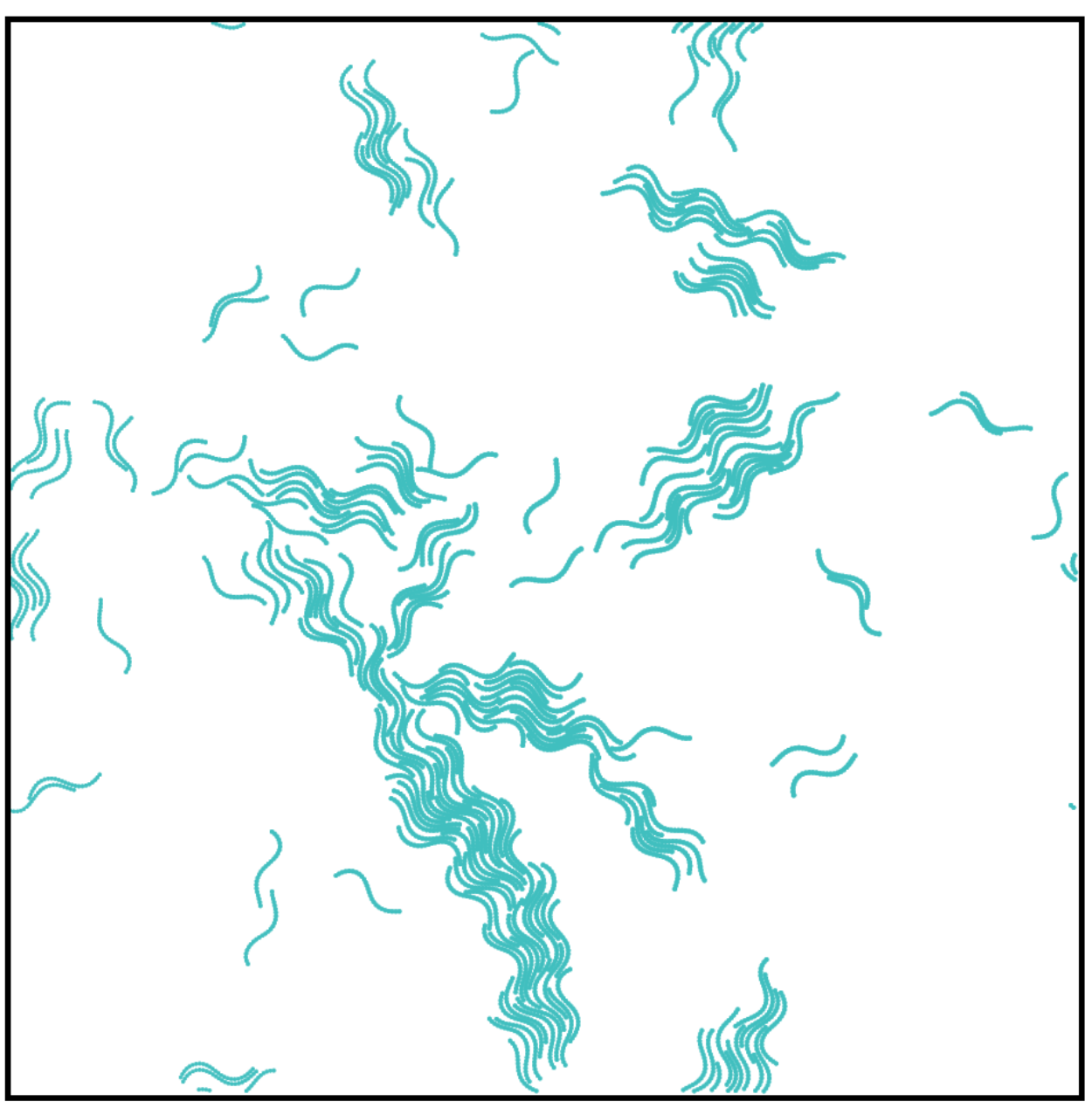}
\caption{(Color online) Snapshot of a multi-flagellum system
in a MPC fluid. The parameters are $\rho_{fl}L_{fl}^2=1.5625$ and
$\sigma_{fl}=0.1\%$. The black box shows the boundary of the
simulation box. Periodic boundary conditions are employed.
For a movie see Ref.~\cite{movie}.}
\label{snapshotflagella}
\end{figure}

Multi-flagellum systems show a similar swarming behavior, consisting of
aggregation and clustering, as observed in Sec.~\ref{sec:swarming_rods}
for self-propelled rods
(see Fig.~\ref{snapshotflagella} and movie~\cite{movie}).
Synchronization of the flagellar beat, and attraction and alignment of
flagella do not only arise from
volume exclusion, as in the SPR systems, but are also triggered by
the hydrodynamic interactions between the sinusoidally undulating
bodies \cite{Yang2008,Elfring2009}.  At the same time, hydrodynamic
interactions between flagella act as a source of environmental noise, which
causes the flagella trajectories to fluctuate strongly.

\subsection{Hydrodynamic Synchronization, Attraction, and Aggregation}
\label{sec:flagella_synchro}

\begin{figure}
\includegraphics[width=8cm]{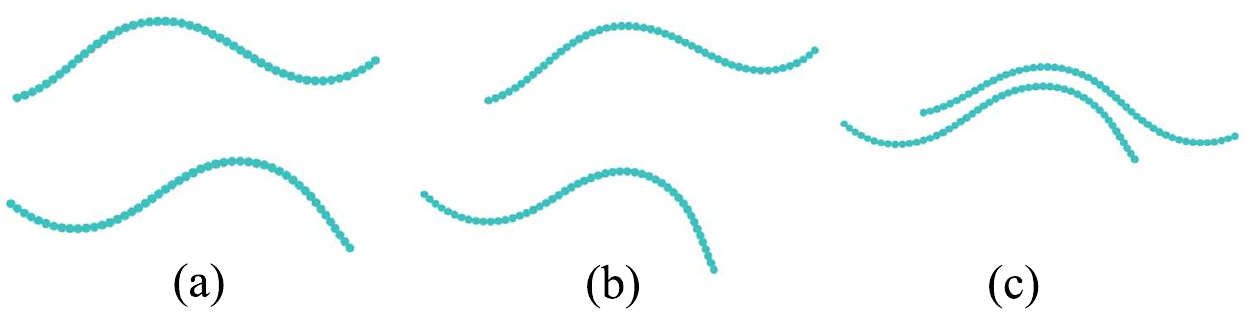}
\caption{(Color online) Synchronization and attraction of
two flagella. The flagella have the same beating frequency $f=1/120$,
and a phase difference $\Delta\varphi=0.5\pi$.
The snapshots are taken at the times (a) $tf=0.167$,
(b) $tf=4.17$ and (c) $tf=22.2$.}
\label{synchronization}
\end{figure}

\begin{figure}
\subfigure{\resizebox{7.0cm}{!}{\includegraphics{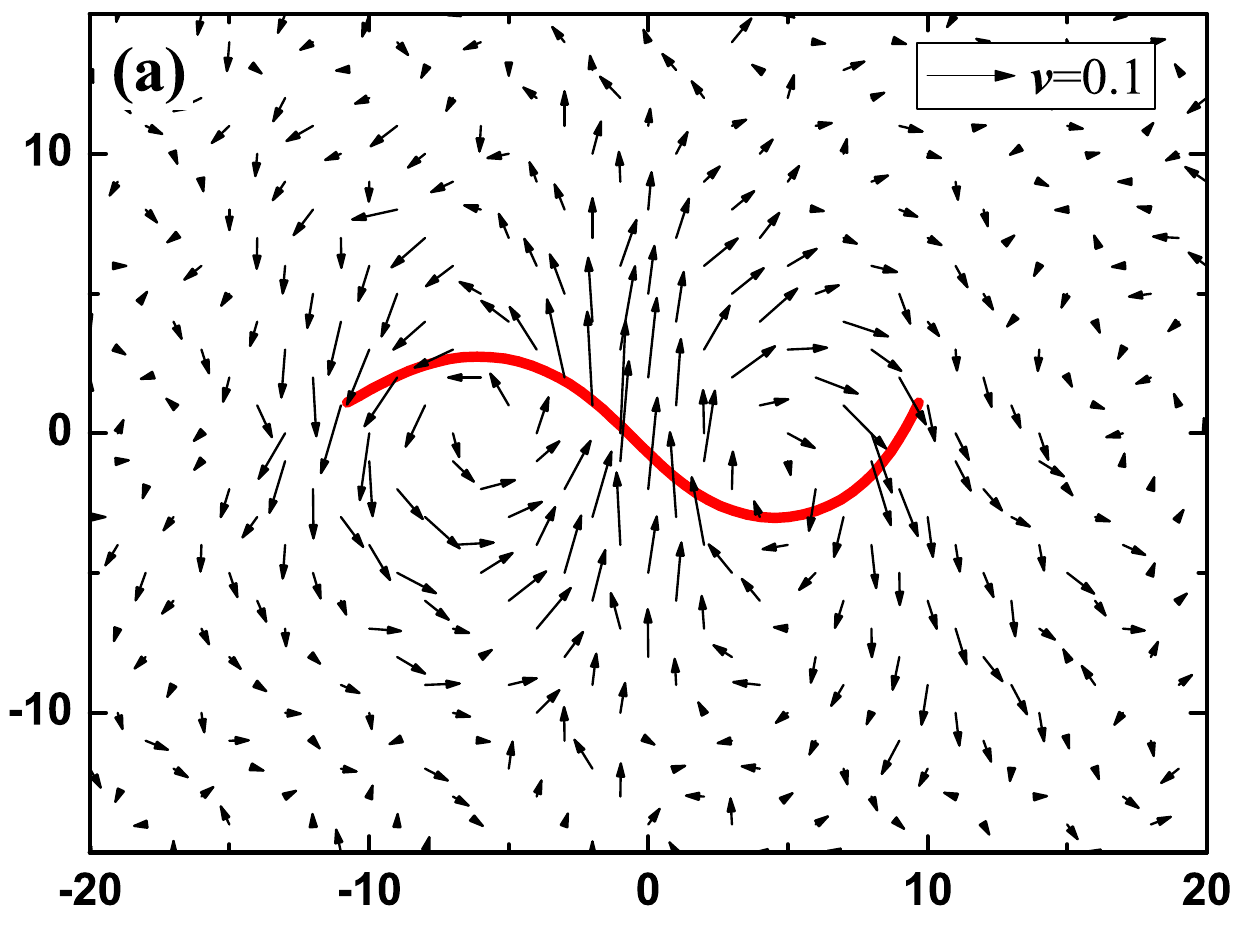}}}
\subfigure{\resizebox{7.0cm}{!}{\includegraphics{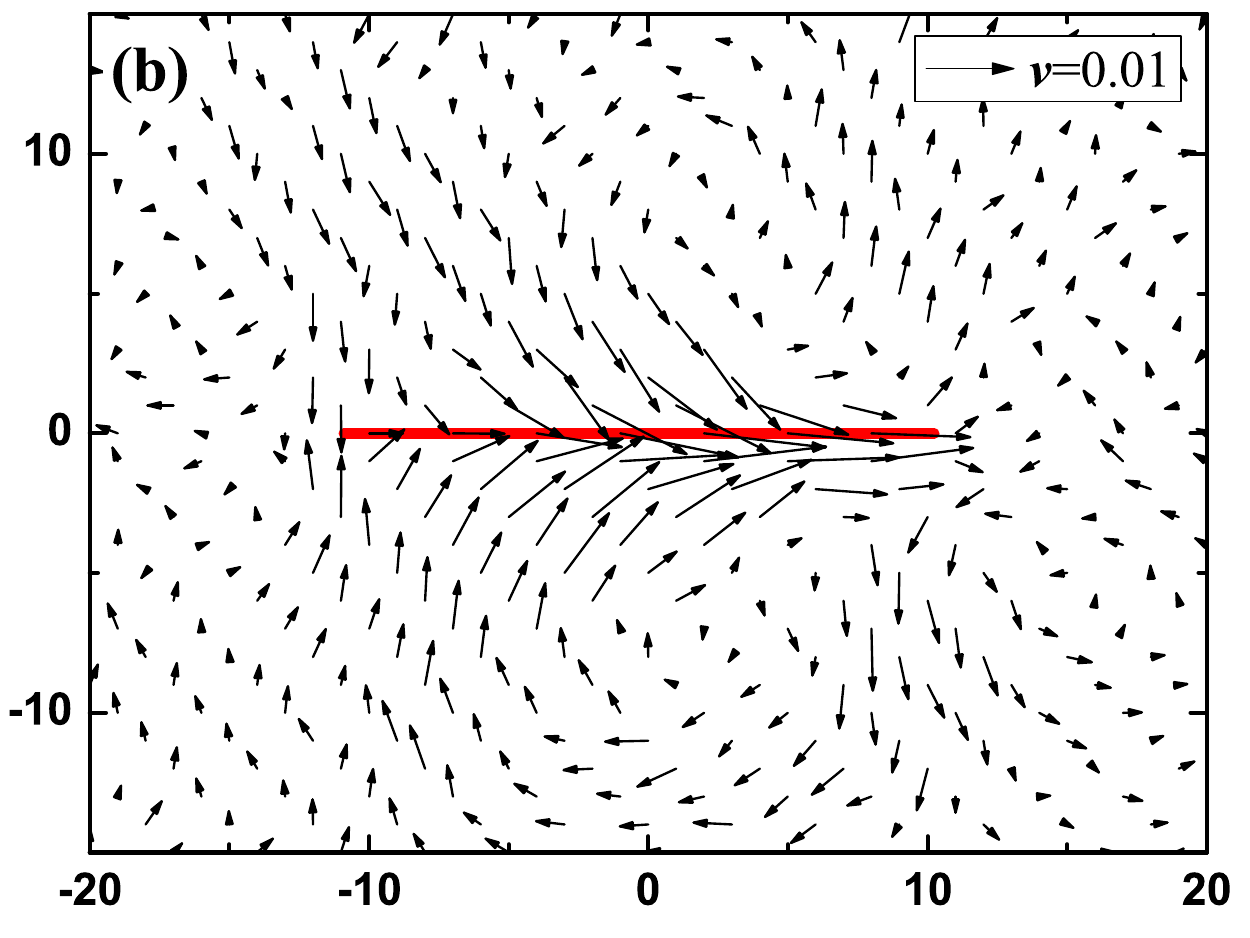}}}
\caption{(Color online) Flow field of a single flagellum, (a) at a
fixed time in the beating cycle, and (b) averaged over the whole
beating cycle. A snapshot of the flagellum and the average flagellum
shape are superimposed in (a) and (b), respectively. The scale bar
indicates the magnitude of the flow velocities. } \label{fig:flow}
\end{figure}

The synchronization and attraction of two flagella is
shown in Fig.~\ref{synchronization}. Synchronization is achieved
within about four beats, while the formation of a tight pair from an initial
distance of about one-third of the flagellar length takes about 20
beats. The flow field of a flagellum is shown in
Fig.~\ref{fig:flow}. The flow field at a certain time in the beating
cycle (Fig.~\ref{fig:flow}a) shows that formation of two vortices,
which propagate from the front to the rear end as the flagellum
moves forward.

The hydrodynamic interaction of swimmers depends on the type of
self-propulsion. The average flow field of flagellum, integrated
over the whole beating cycle, demonstrates that the flagellum, which
might be expected to be a ``neutral" swimmer ({\it i.e.}, neither a
pusher nor a puller) is indeed a very weak pusher
--- where the dominant propulsion is located closer to the rear end
--- because the line connecting the centers of the two vortices
intersects the average flagellum shape behind its mid-point
(Fig.~\ref{fig:flow}b). This generates a in-flow from both sides of
the flagellum near the front end, which is responsible for
hydrodynamic attraction \cite{Ishikawa2009,Lauga2009}.

In multi-flagellum systems, large clusters can form by collisions
of smaller clusters, supported by the hydrodynamic attraction between
neighboring flagella; large clusters can disintegrate
into smaller components due to the diversity of flagellar
frequencies, or the hydrodynamic flow fields of other clusters.
With hydrodynamic
interactions, large clusters of flagella are usually strongly
extended in their direction of motion, as shown in
Fig.~\ref{snapshotflagella} and movie~\cite{movie}.
The flagella inside the cluster are well synchronized. This
structure is reminiscent of the ``sperm-train" structure observed in
rodent-sperm experiments \cite{Moore2002,Immler2007}. The
elongated clusters can extend to distances as large as the side
length of the simulation box, which induces strong finite-size
effects.

Similar to the definition of a rod cluster in
Sec.~\ref{sec:swarming_rods}, a flagellum cluster is
defined as a set of flagella that are connected or neighbors either
directly or through other agents at a given moment in time. Its size
is the number $n$ of flagella it contains. A freely-swimming single
flagellum is considered as a cluster of size $n=1$.

\subsection{Cluster-Size Distributions}

\begin{figure}
\includegraphics[width=7.5cm]{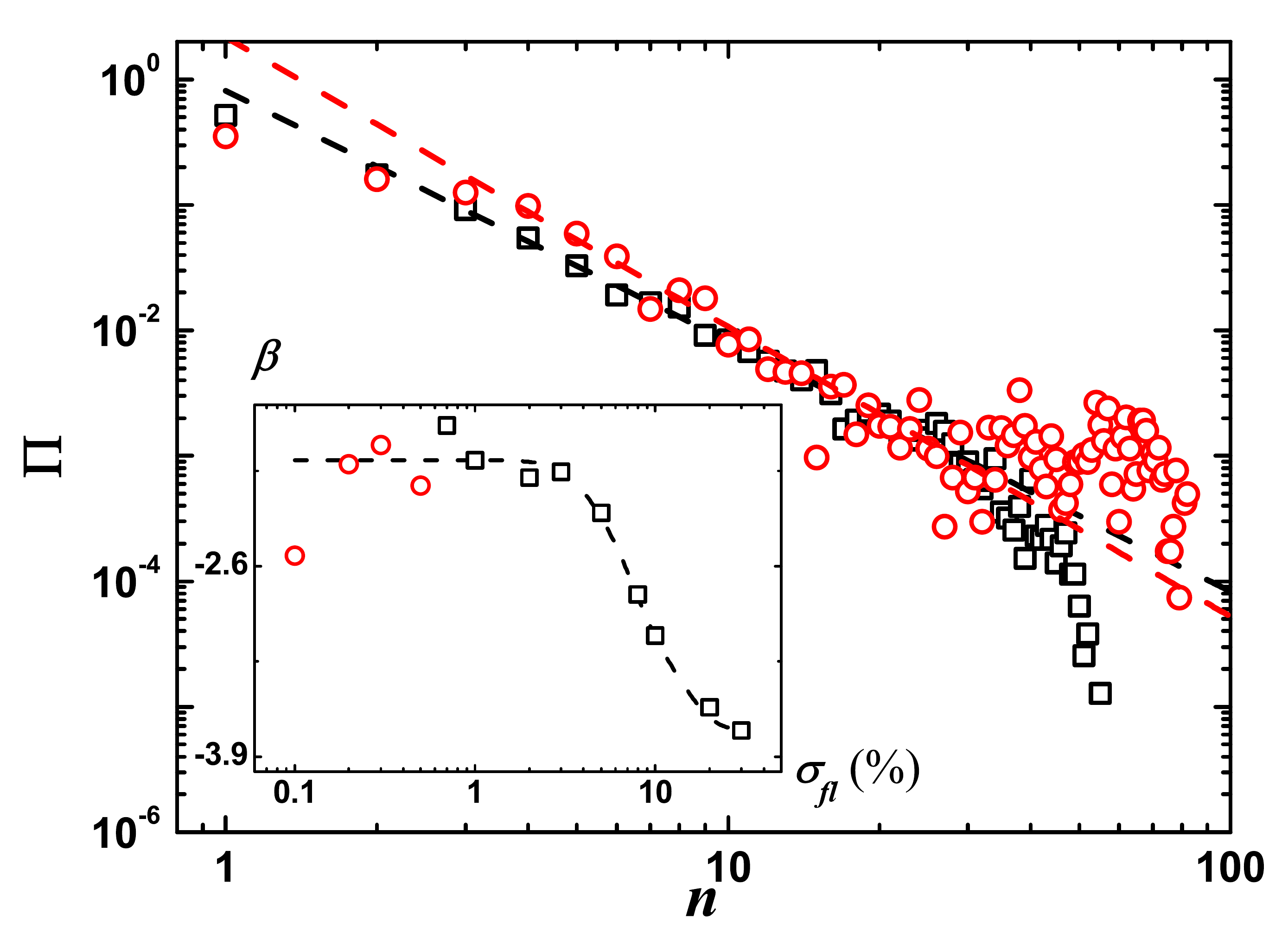}
\caption{(Color online) The two different types of cluster-size
probability density functions
$\Pi_1(n)$ with $\sigma_{fl}=2\%$ ($\square$, black),
and $\Pi_2(n)$ with $\sigma_{fl}=0.1\%$ ($\bigcirc$, red),
observed in multi-flagellum systems with density
$\rho_{fl}L_{fl}^2=1.5625$.
The dashed lines are fits to the power-law parts of each PDF.
The inset shows the exponent $\beta$ as a function of the
variance $\sigma_{fl}$ of the frequency distribution.
Symbols indicate the systems characterized by
$\Pi_2$ ($\bigcirc$, red) and $\Pi_1$ ($\square$, black). The dashed
line is a guide to the eye.}
\label{Piflagellum}
\end{figure}

Both probability density functions $\Pi_1$ and $\Pi_2$ are
observed in our multi-flagellum
systems, as shown in Fig.~\ref{Piflagellum}. The variance
$\sigma_{fl}$ of the distribution of beat frequencies
is used as a measure of
the noise level. At low $\rho_{fl}$ or high
$\sigma_{fl}$, we find $\Pi_1$; at high $\rho_{fl}$ or low
$\sigma_{fl}$, we observe $\Pi_2$.
In contrast to $\Pi_2$ for SPR systems in Sec.~\ref{sec:PDF_rods},
$\Pi_2$ for flagella systems displays a deviation from
the power-law behavior for very small cluster sizes, $n=1$ and $n=2$.
We believe that this is due to the hydrodynamic synchronization
and attraction of neighboring flagella.
For flagella, we have never observed a giant cluster with a
blocked structure, in contrast to the SPR system of Fig.~\ref{snapshot}c.

Although the distribution of beating frequencies is an internal
property of the swimmers, the influence of $\sigma_{fl}$ on the
exponent $\beta$ of Eq.~(\ref{eq:PDF_powerlaw})
is similar to the influence of the environmental noise in our
previous SPR simulations, as shown in the inset of Fig.~\ref{Piflagellum}.
$\beta$ is nearly constant for $\sigma_{fl}<3\%$, then decreases
smoothly with increasing $\sigma_{fl}$.

\begin{figure}
\includegraphics[width=7cm]{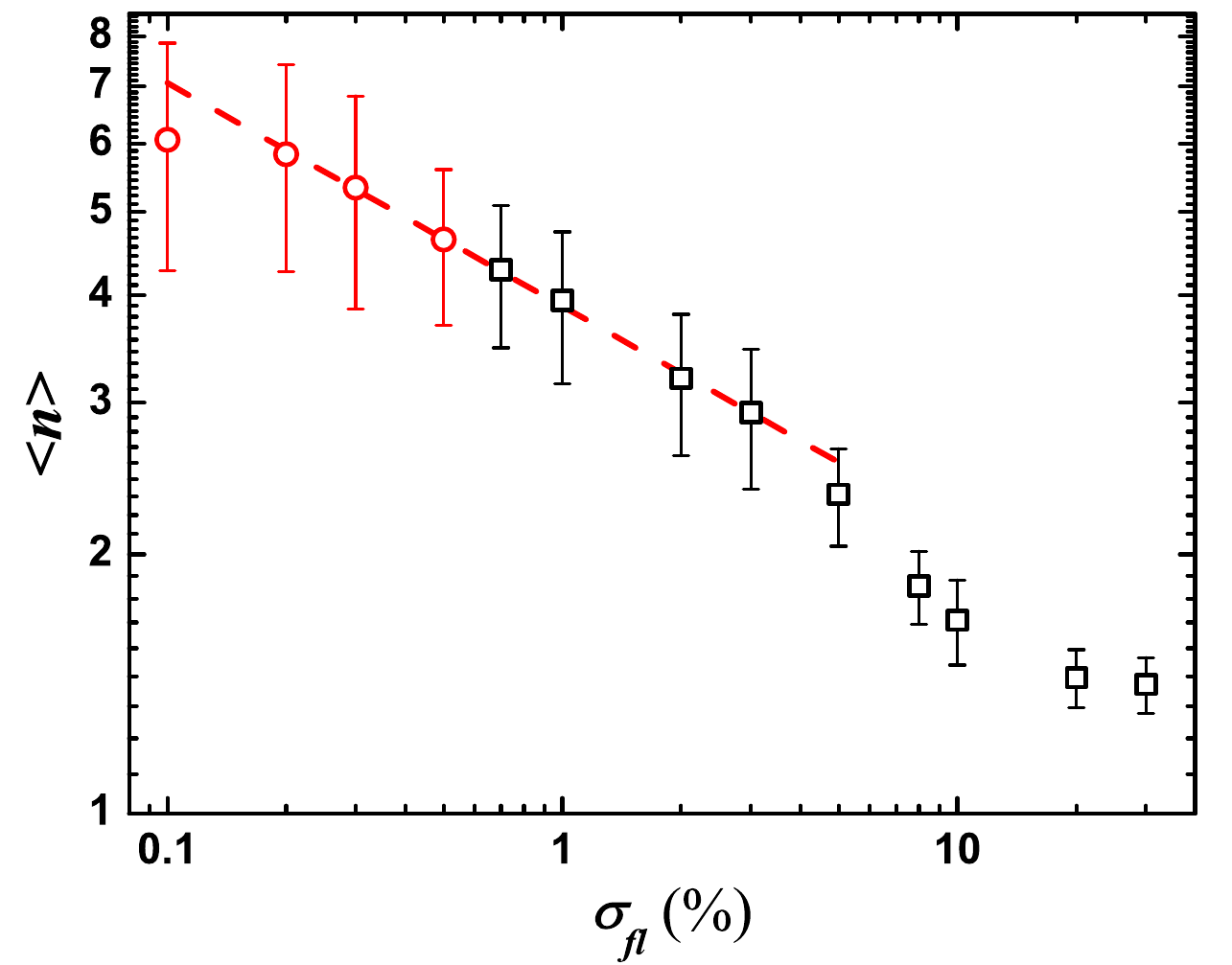}
\caption{(Color online) Stationary average cluster size $\langle
n\rangle$ versus the variance $\sigma_{fl}$ of the frequency
distribution, for flagellar density
$\rho_{fl}L_{fl}^2=1.5625$. Symbols indicate systems
with $\Pi_1$ ($\square$, black) and $\Pi_2$ ($\bigcirc$, red). The
error bars are the standard deviation of the fluctuations. The
dashed line indicates a power-law decay with an exponent $\zeta=0.26$. }
\label{averagen}
\end{figure}

The average cluster size $\langle n\rangle$ in the stationary state
is a function of $\sigma_{fl}$, as shown in Fig.~\ref{averagen}.
Increasing $\sigma_{fl}$ results in an increase of the overall
break-up rate; hence $\langle n\rangle$ decreases. In the large
$\sigma_{fl}$ limit, $\langle n\rangle$ approaches unity,
corresponding to a disordered state with randomly distributed
flagella.
The power-law decay
\begin{equation}
\langle n \rangle \sim \sigma_{fl}^{-\zeta}
\end{equation}
of the average cluster size
with exponent $\zeta\simeq 0.26$ emphasizes the universality
of the swarming behavior of rSPP systems in two dimensions. The power-law
scaling of $\langle n\rangle$ as a function of $\sigma_{fl}$
implies a divergence when $\sigma_{fl}\rightarrow 0$.
We believe that the small deviation from
the power-law behavior for $\sigma_{fl}=0.1\%$ in
Fig.~\ref{averagen}, as well as the deviation of $\beta$
from the plateau value for $\sigma_{fl}=0.1\%$ in
Fig.~\ref{Piflagellum}, are due to finite-size effects.

\begin{figure}
\includegraphics[width=7.4cm]{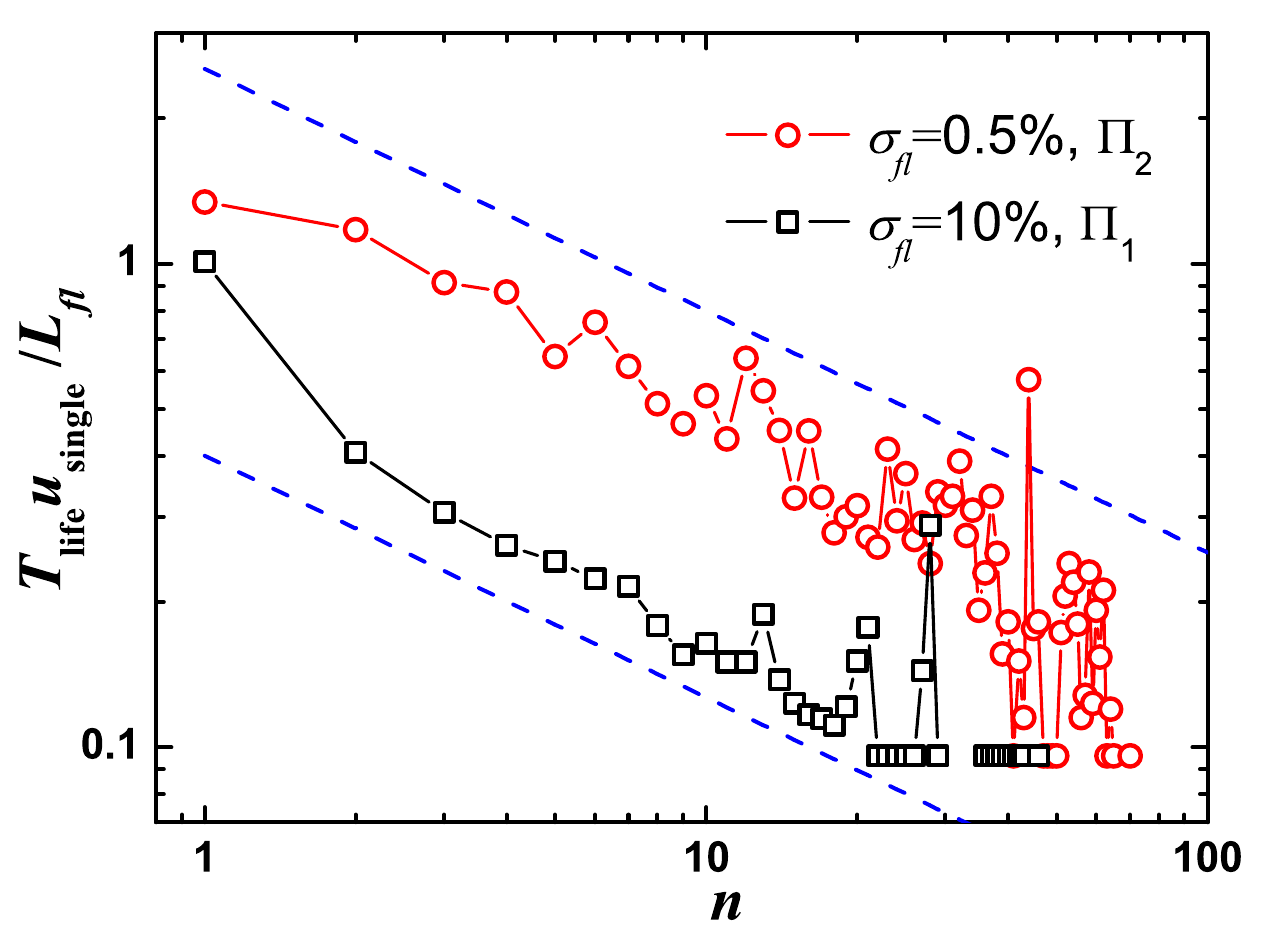}
\caption{(Color online) The lifetime
$T_{life}v_{single}/L_{fl}$ of flagella clusters versus their size
$n$. The flagella number density is $\rho_{fl}L_{fl}^2=1.5625$.
The dashed lines indicate a power law with exponent $\delta=0.5$.}
\label{fig:lifetimeflagellum}
\end{figure}

\subsection{Cluster Lifetimes}

The average cluster lifetime $T_{life}(n)$ decreases as an
effective power-law function of cluster size $n$, see
Eq.~(\ref{eq:lifetime}), with an exponent $\delta\simeq 0.5$,
as shown in Fig.~\ref{fig:lifetimeflagellum}. The value of $\delta$ is
very close to the prediction based on the assumption of a mechanism
of particle accumulation and shedding proportional to the cluster
perimeter, as presented in Sec.~\ref{sec:lifetime_rods}. This good
agreement provides further evidence for the different mechanisms
of cluster stabilization for rods and flagella, which are a (partially)
blocked motion and a hydrodynamic attraction, respectively.

Note that the system size of the flagella simulations is not as large
as for the SPR systems.  Thus, the effective power law can only be
observed over a smaller range of cluster sizes.
In SPR simulations, single rods ($n=1$)
always have a much longer lifetime compared to expectation from the
effective power law, see Fig.~\ref{fig:lifetimevssigma}. In contrast, for
flagella with full hydrodynamic interactions, $T_{life}(1)$ is much
closer to the power-law extrapolation, and can even be lower
than the power-law prediction ({\em e.g.} for $\sigma_{fl}=0.5\%$ in
Fig.~\ref{fig:lifetimeflagellum}).

\subsection{Comparison of Sperm and Flagella}

As explained in Sec.~\ref{sec:model_flagella}, our model of a
flagellum differs from the model of a sperm employed in
Ref.~\cite{Yang2008} by the absence of a passive midpiece and a
circular head. Also, in the sperm simulations \cite{Yang2008}, two
sine waves were present on the tail, while a single sine wave is
present on the flagellum.

\begin{figure}
\includegraphics[width=6.8cm]{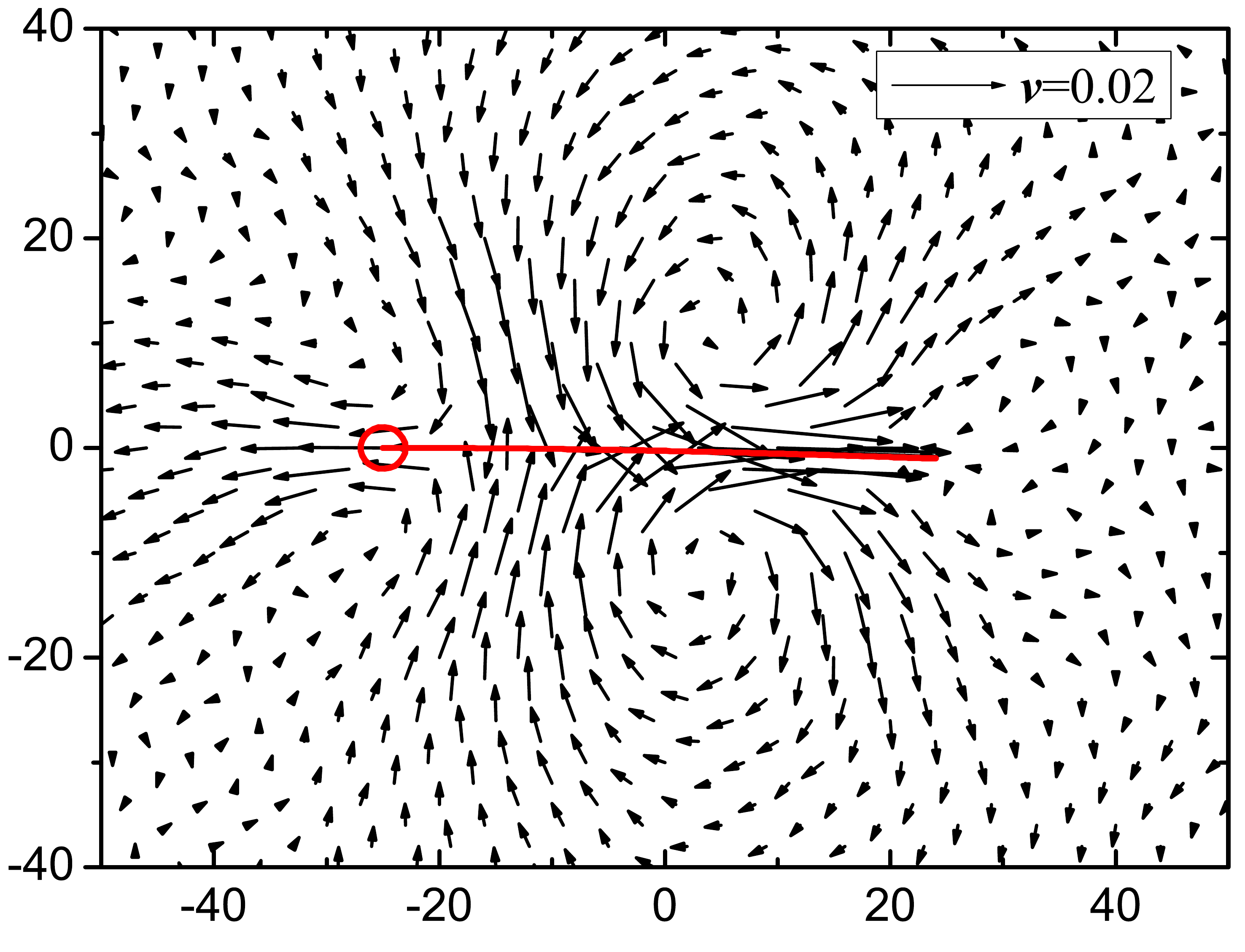}
\caption{(Color online) Flow field of a single sperm, beating
with two sine waves on its tail, averaged over the whole beating cycle.
The average sperm shape is superimposed. The scale bar indicates
the magnitude of the flow velocities. }
\label{fig:sperm}
\end{figure}

How similar or different is the collective behavior of sperm and
flagella? There are three different aspects to this question.
{\em Synchronisation} depends mainly on the interaction of the
time-dependent oscillatory flow field of two neighboring flagella
\cite{Taylor1951,Elfring2009},
and is therefore very similar, as can be seen from the results
presented in Sec.~\ref{sec:flagella_synchro} and those of
Ref.~\cite{Yang2008}.
On the other hand, the {\em hydrodynamic attraction} of sperm and
flagella is quite different. A sperm cell,
consisting of a flagellum and a large head, is clearly a pusher,
as demonstrated by the average flow field of a sperm in
Fig.~\ref{fig:sperm}. The flagellum pushes the fluid backward in
both cases, but the bulky head of the sperm drags the fluid forward
much more strongly, which generates the characteristic sidewise
inflow of fluid towards the midpiece region
\cite{Lauga2009,Ishikawa2009,Elgeti2010}.
In contrast, flagella are very weak pushers, as demonstrated in
Fig.~\ref{fig:flow}b above.
Therefore, sperm have stronger hydrodynamic attraction than
flagella.

Finally, the {\em swarming behavior} in both flagella and sperm system
is characterized by cluster-size distributions and the dependence of
the average cluster size on the width $\sigma_{fl}$ of the distribution
of beat frequencies. While the cluster-size distribution of flagella
follows a power-law decay over a wide range, it was not possible to
clearly identify a power-law behavior for sperm in Ref.~\cite{Yang2008}
due to the relatively small systems of 25 and 50 sperm.
The average cluster size is found to depend on $\sigma_{fl}$ as
$\langle n \rangle \sim \sigma_{fl}^{-\zeta}$, with $\zeta=0.20$ for
sperm \cite{Yang2008} and $\zeta=0.26$ for flagella.
Larger systems have to be investigated to
see whether the exponents $\zeta$ for sperm and flagella are different
or not. In any case,
the stronger hydrodynamic attraction of sperm, which favors larger
cluster sizes, is partially offset by the bulky head of sperm, which
implies that  the sperm clusters in Ref.~\cite{Yang2008} are much
more loosely packed than the flagella clusters studied here.

%
%

\section{Summary and Conclusions}
\label{sec:conclusions}

We have simulated systems of rigid rods propelled
by a constant force along their long axis, and systems of flagella
propelled by a sinusoidal beating motion, in two dimensions.
In both systems, we observe cluster formation and break-up,
controlled by the particle density and the internal or external noise.
In our simulations, the particle density is always much
lower than the critical density of a nematic phase in thermal equilibrium.

Without any attractive potential, self-propelled rods (SPR) exhibit
an aggregation behavior triggered only by volume exclusion. Three
characteristic types of cluster-size probability density functions
$\Pi(n)$ appear in different regions of a dynamic phase diagram of
stationary states. At high noise and low density, the system is
characterized by $\Pi_1$, which shows a power-law distribution over
a range of cluster sizes, with an exponential cutoff at large
cluster sizes. At low noise and high density, the system is in a
state characterized by $\Pi_3$, which has a peak at sizes near the
total number of particles in the system, representing a giant
cluster. Systems in an intermediate region of noise and density are
characterized by $\Pi_2$, which is a transition state between
$\Pi_1$ and $\Pi_3$. It has a bimodal shape, with a power-law decay
at small cluster sizes and a shoulder at larger sizes. Clusters in
$\Pi_1$ and $\Pi_2$ systems retain a high motility,
whereas the giant clusters found in the third state is almost
immobile due to its blocked configuration.
The average cluster size at equilibrium, directly related to
cluster-size distribution $\Pi$, displays a power-law dependence
with decreasing noise amplitude before the system reaches the
$\Pi_3$ state.

Sinusoidally beating flagella were simulated in a low-Reynolds-number
fluid with full hydrodynamics as an example of self-propelled
rod-like particles with explicit propulsion mechanism.
Flagella synchronize their beats and attract
each other through the hydrodynamic interactions. Despite the
different propulsion mechanisms, the basic swarm behavior of
aggregation and clustering observed for swimming flagella is
remarkably similar to the behavior seen in SPR systems. We
observe both $\Pi_1$ and $\Pi_2$ cluster-size probability density
functions by varying the width $\sigma_{fl}$ of the flagellar
beat-frequency distribution, which acts as a source of internal noise
in the system.
The average cluster size also display a power-law dependence on
$\sigma_{fl}$, as for SPR systems.

Despite these similarities in the clustering behavior, the two systems
show some important differences. They can be traced back to the
hydrodynamic attraction between beating flagella, which is absent
in our simulations of self-propelled rods.
First, the configurations of the flagella clusters
consist of tightly stacked flagella with synchronized
shapes, and extend in their moving directions. Those elongated
clusters are reminiscent of the huge, mobile ``sperm trains" observed
in rodent-sperm experiments \cite{Moore2002}.
Clusters in the SPR systems are more compact, and have a wedge-like
structure, which arises from the partially blocked rod motion responsible
for the cluster aggregation, as well as from collisions with other
clusters.
Second, the $\Pi_3$ state of a completely blocked structure, which is
observed for SPR at high density and low noise, does not seem to
exist in flagellar systems.
Third, the cluster lifetimes decay with different effective power
laws, $\delta=0.2$ for SPR and $\delta=0.5$ for flagella.
Finally, hydrodynamic interactions between different flagella clusters
act as an additional source of noise and contribute to increase the
break-up rate.

The existence of the giant, immobile cluster should
depend sensitively on the aspect ratio and the type and range of the
interactions between self-propelled rods, where longer rods and
shorter-range interaction favors the giant-cluster formation. This
conclusion follows from the result of Ref.~\cite{Peruani2006} for
rods of aspect ratio $L_{rod}/l_b \le 12$ that $\Pi_1$-$\Pi_2$
boundary shifts to higher density with decreasing rod length, and
our result of Fig.~\ref{averclustersize} that the $\Pi_2$ state
corresponds to ``supercooled" liquid state which transforms into
the $\Pi_3$ state once a giant-cluster nucleus has formed.
Blocked clusters were not seen in Ref.~\cite{Peruani2006} for
rod lengths $L_{rod}/l_b \le 12$
due to the relatively small system size with $N_{rod}=100$. However,
blocked states were observed in Ref.~\cite{Kraikivski2006}
for a much larger rod length, $L_{rod}/l_b=40$, already for
a system of only about 50 rods at density $\rho_{rod}L_{rod}^2=2$.

Our simulations have been restricted to the
isotropic phase of rods in thermal equilibrium. It will be
interesting to see in the future whether immobile, blocked states
can also exist (or even dominate) in the nematic regime, or whether
they are suppressed by the preferred rod orientation.

In the light of our results, we conclude that different systems of
rod-like self-propelled
particles display a universal swarming behavior, but also specific
properties related to their propulsion mechanisms and the
presence or absence of hydrodynamic interactions.

\acknowledgments
We thank Jens Elgeti and Roland Winkler for stimulating discussions.
Yingzi Yang acknowledges support by the International
Helmholtz Research School on Biophysics and Soft Matter (IHRS BioSoft).
Vincent Marceau is grateful to the RISE program of the DAAD (Germany)
and to NSERC (Canada) for financial support. This work was supported in
part by the VW foundation through the program ``Computational Soft Matter
and Biophysics".


\end{document}